\documentclass[aps,twocolumn,groupedaddress,superscriptaddress,floatfix,amsmath,amssymb,prb]{revtex4-2}
\usepackage{amsmath}
\usepackage{amssymb}
\usepackage{graphicx}
\usepackage{textcomp}
\usepackage{bm}

\usepackage{booktabs}
\usepackage{siunitx}
\usepackage{tabularx}
\usepackage{multirow}
\usepackage{dsfont}
\usepackage{braket}
\usepackage{bbm}
\usepackage{bbold}

\usepackage[usenames,dvipsnames]{color}	

\usepackage[colorlinks,citecolor=Blue,linkcolor=Red, urlcolor=Blue]{hyperref}

\usepackage{graphicx} 
\usepackage{float}
\usepackage{fancyhdr}
\usepackage{fnpos}
\usepackage[english]{babel}
\addto{\captionsenglish}{%
  
}
\usepackage{array}
\usepackage{droidsans}
\usepackage{charter}
\usepackage[T1]{fontenc}
\usepackage[usenames,dvipsnames]{xcolor}
\usepackage{setspace}
\usepackage[compact]{titlesec}
\usepackage{hyperref}

\DeclareUnicodeCharacter{2212}{-}

\begin{document}

\title{Controlled coherent  dynamics of [VO(TPP)], a prototype molecular nuclear qudit with an electronic ancilla}

\author{Simone Chicco}
\affiliation{\textit{~Universit\`a di Parma, Dipartimento di Scienze Matematiche, Fisiche e Informatiche, I-43124 Parma, Italy.}}
\affiliation{\textit{~UdR Parma, INSTM, I-43124 Parma, Italy.}}
\author{Alessandro Chiesa}
\affiliation{\textit{~Universit\`a di Parma, Dipartimento di Scienze Matematiche, Fisiche e Informatiche, I-43124 Parma, Italy.}}
\affiliation{\textit{~UdR Parma, INSTM, I-43124 Parma, Italy.}}
\author{Giuseppe Allodi}
\affiliation{\textit{~Universit\`a di Parma, Dipartimento di Scienze Matematiche, Fisiche e Informatiche, I-43124 Parma, Italy.}}
\author{Elena Garlatti}
\affiliation{\textit{~Universit\`a di Parma, Dipartimento di Scienze Matematiche, Fisiche e Informatiche, I-43124 Parma, Italy.}}
\affiliation{\textit{~UdR Parma, INSTM, I-43124 Parma, Italy.}}
\author{Matteo Atzori}
\affiliation{\textit{~Dipartimento di Chimica “Ugo Schiff” \& INSTM, Universit\`a Degli Studi di Firenze, I-50019 Sesto Fiorentino, Italy.}}
\affiliation{\textit{~Laboratoire National des Champs Magnétiques Intenses (LNCMI), Univ. Grenoble Alpes, INSA Toulouse, Univ. Toulouse Paul Sabatier, EMFL, CNRS, F-38043 Grenoble, France.}}
\author{Lorenzo Sorace}
\affiliation{\textit{~Dipartimento di Chimica “Ugo Schiff” \& INSTM, Universit\`a Degli Studi di Firenze, I-50019 Sesto Fiorentino, Italy.}}
\author{Roberto De Renzi} 
\affiliation{\textit{~Universit\`a di Parma, Dipartimento di Scienze Matematiche, Fisiche e Informatiche, I-43124 Parma, Italy.}}
\author{Roberta Sessoli}
\affiliation{\textit{~Dipartimento di Chimica “Ugo Schiff” \& INSTM, Universit\`a Degli Studi di Firenze, I-50019 Sesto Fiorentino, Italy.}}
\author{Stefano Carretta}
\affiliation{\textit{~Universit\`a di Parma, Dipartimento di Scienze Matematiche, Fisiche e Informatiche, I-43124 Parma, Italy.}}
\affiliation{\textit{~UdR Parma, INSTM, I-43124 Parma, Italy.}}

\date{\today}
\email{E-mail: stefano.carretta@unipr.it}

\begin{abstract} 
We show that [VO(TPP)] (vanadyl tetraphenylporphyrinate) is a promising system suitable to implement quantum computation algorithms based on encoding information in multi-level (qudit) units. Indeed, it embeds an electronic spin 1/2 coupled through hyperfine interaction to a nuclear spin 7/2, both characterized by remarkable coherence. We demonstrate this by an extensive broadband nuclear magnetic resonance study, which allow us to characterize the nuclear spin-Hamiltonian and to measure the spin dephasing time as a function of the magnetic field. In addition, we combine targeted measurements and numerical simulations to show that nuclear spin transitions conditioned by the state of the electronic qubit
can be individually addressed and coherently manipulated by resonant radio-frequency pulses, thanks to 
the remarkably long coherence times and the effective quadrupolar coupling induced by the strong hyperfine coupling. This approach may open new perspectives for developing new molecular qubit-qudit systems.
\end{abstract}

\maketitle

\footnotetext{\dag~Electronic Supplementary Information (ESI) available: [details of any supplementary information available should be included here]. See DOI: 00.0000/00000000.}

\section{Introduction}
The physical implementation of quantum information processing has known an astonishing boost in the last few years, with huge efforts devoted to realize devices able to solve problems impossible for a classical machine \cite{Supremacy}.
Most of the proposed architectures are based on encoding elementary units of information in two-level quantum systems called {\it qubits}, in perfect analogy with classical digital computation.
In spite of great progresses in this direction, with important results already achieved by leading technologies \cite{RevTacchino,NatPhysIBM,HFGoogle,D0SC01908A,vqe1,RevTrappedIons},
current realizations are still noisy and only allow one to perform approximate small-size calculations \cite{NISQ}. \\
Hence, it is crucial to explore different approaches and new  
physical systems, in order to reduce the complexity of manipulations, thus making important results achievable in the short term, even with noisy devices.
This could be obtained by replacing the elementary two-level building blocks with multi-level ({\it qudit}) systems, which provide a powerful alternative to the conventional encoding \cite{RevQudits,Imany,PhysRevA.91.042312,PhysRevA.87.022341}.
Indeed, qudits are characterized by a larger Hilbert space than simple qubits, which can be exploited to store, process and transfer quantum information. This yields in many cases a significant reduction of the number of units and operations required to implement an algorithm \cite{RevQudits}.
In particular, working with larger Hilbert spaces reduces the number of units needed to represent an arbitrary unitary matrix. Hence, some fundamental gates such as the Toffoli gate \cite{NatPhysToffoli} or algorithms such as Deutsch \cite{DeutschQudit}, Grover \cite{GroverWW}, Quantum Fourier Transform, or Quantum Phase Estimation can be implemented much faster and using fewer operations \cite{RevQudits}.
Quantum simulation schemes have recently been put forward, exploiting the qudits multi-level structure to represent bosonic fields and to mimic light-matter interactions \cite{TacchinoQudits}.\\
In addition, qudit-based architectures are more resistant to error sources acting locally on each unit, thus reducing the effect of noise and making this implementation very appealing in the short-term \cite{RevQudits}. More importantly, the extra space of the qudit can be used to encode qubits with embedded quantum-error correction, a fundamental step to make the quantum hardware resistant to environmental noise \cite{PRXGirvin,Mancini,PRXPreskill,jacsYb,JPCLqec,ErCeEr}. \\
A natural platform to implement qudit-based logic is provided by Molecular Nanomagnets (MNMs).
Indeed, the low-energy levels of these molecules can be 
easily accessible through microwave or radio-frequency pulses and often characterized by long coherence times \cite{Gaitarev,Sessoli2019}. By exploiting the power of coordination chemistry, the multi-level structure needed to encode a qudit can be obtained, for example, by combining strongly-interacting magnetic centers, yielding a ground state with a sizable total spin \cite{Mn19powell,giant}. 
The high-degree of chemical tunability of these magnetic molecules allows also devising synthetic strategies to improve their coherence times \cite{PRLWedge,Bader,Zadrozny,Atzori2016,Atzori_JACS,Atzori2017,Atzori2018,Freedman_JACS,Freedman_Cr,Freedman2014,Zadrozny_Fe} and to link complexes together in order to fit specific requirements of quantum computation schemes \cite{Luis2011,Aguila2014,Ardavan2015,SciRepNi,modules,Chem,VO2,Entanglement,Heterodimers}. \\
Remarkably, the simplest chemical realization of a $2I+1$ levels qudit is provided by a single magnetic-ion complex, embedding 
a nuclear \cite{GroverWW,jacsYb,JPCLqec} spin $I$ hyperfine coupled to an electronic spin doublet. Such a magnetic coupling can be exploited for the quantum simulation of physical models involving many degrees of freedom (such as photons interacting with matter \cite{TacchinoQudits}) or to implement quantum error-correction codes using the electronic qubit as an ancilla for error detection \cite{jacsYb,JPCLqec}. In addition, the hyperfine coupling of the nuclear qudit with an electronic spin can significantly speed-up qudit manipulations \cite{Actuator,jacsYb} by inducing a mixing between electronic and nuclear spin states. Such a mixing, yielding level anti-crossing at low magnetic field, could also be exploited to enhance sensitivity in quantum metrology applications \cite{PRBTroiani2016,PRAGhirardi2018}. \\
Here we show that the [VO(TPP)] (TPP = tetraphenylporphyrin) molecular complex ({\bf 1}) is a prototypical qubit-qudit system to realize such a platform. Indeed, it provides a large spin ($I=7/2$, eight levels) nuclear qudit, coupled by hyperfine interaction to an electronic spin 1/2, both characterized by remarkable coherence. 
The electronic coherence and spin Hamiltonian were already characterized in previous works \cite{VOTPP,Bonizzoni} and different electronic transitions (depending on the nuclear spin state) were recently assessed by broadband electron paramagnetic resonance \cite{gimeno2021broadband}.
However, to exploit the nuclear qudit in a quantum information perspective, a deeper understanding of the nuclear degrees of freedom is mandatory.
We perform here a thorough experimental study on single crystals of {\bf 1} by broadband nuclear magnetic resonance (NMR). By measuring NMR spectra on a targeted frequency and magnetic field range, we characterize in detail the nuclear spin Hamiltonian, including small quadrupolar interactions, and the hyperfine coupling with the electronic qubit. 
We then address single-quantum coherences and study Hahn echo decays as a function of the external field, finding remarkably long phase memory times. 
In spite of the small quadrupole interaction typical of $^{51}$V nucleus, we show that hyperfine coupling yields an effective quadrupolar splitting arising along specific magnetic field directions. 
This allows us to resolve all nuclear transitions, yielding
full control of the system by radio-frequency (rf) pulses resonant with single-quantum transitions, as demonstrated by measurements and simulations of Rabi oscillations. \\
Hence, our study shows that, the class of systems suitable to implement qudit algorithms is not limited to nuclear spins characterized by an intrinsic large quadrupolar coupling, but can be extended, e.g., to V complexes, typically showing very long coherence \cite{Zadrozny,Freedman_JACS,Atzori2016,Phonons}.
The combination of long coherence times and addressability of each transition makes all this class of systems promising to implement a quantum simulator \cite{TacchinoQudits}, a qubit with embedded quantum-error correction \cite{JPCLqec} or more generally for qudit-based quantum logic \cite{RevQudits}.  \\

\section{Results and discussion}

\subsection{Synthesis}
The complex [VO(TPP)] has been prepared as reported in \cite{VOTPP,Synthesis} and it has been magnetically diluted ($2 \%$) in its diamagnetic analogue [TiO(TPP)], prepared with the same method, by cocrystallization. Slow solvent evaporation (CH$_3$COCH$_3$) of the mixture of isostructural starting materials provides single crystals of [TiO(TPP)]:[VO(TPP)] that assume the shape of a square pyramid with well developed ($\pm 1$,$\pm 1$,$\pm 1$) faces and dimensions of ca. $4 \times 4 \times 2$ mm$^3$ (see Figure S1). The system crystallizes in the tetragonal \emph{I}4 space group with the four-fold symmetry axis \emph{c} corresponding to the direction of the V=O bond. A description of the molecular and crystal structure of [VO(TPP)] and [TiO(TPP)] is reported in ref. \cite{VOTPP}. \\

\subsection{Characterization by broadband NMR}
\label{section:HamiltonianFitting}
$^{51}$V ($I=7/2$ nat. abund.$=99.85 \%$) NMR spectra are measured by a broadband home-built NMR spectrometer (see ESI and Ref. \cite{Spectrometer}) at fixed temperature $T=1.4$ K on a single crystal of [VO(TPP)], diluted at $2\%$ in its isostructural diamagnetic host [TiO(TPP)] in order to reduce inter-molecular dipole-dipole interactions. We collect spectra by measuring Hahn-echoes as a function of frequency, for different applied static fields $B_0$ 
in the range $0.05 - 0.3$ T, along the two orthogonal crystal directions, corresponding to the two inequivalent symmetry directions of the [VO(TPP)] molecule: \emph{ab} plane and \emph{c} axis (see SI for further details). \\
\begin{figure}[h]
  \centering
  \includegraphics[width=1\linewidth]{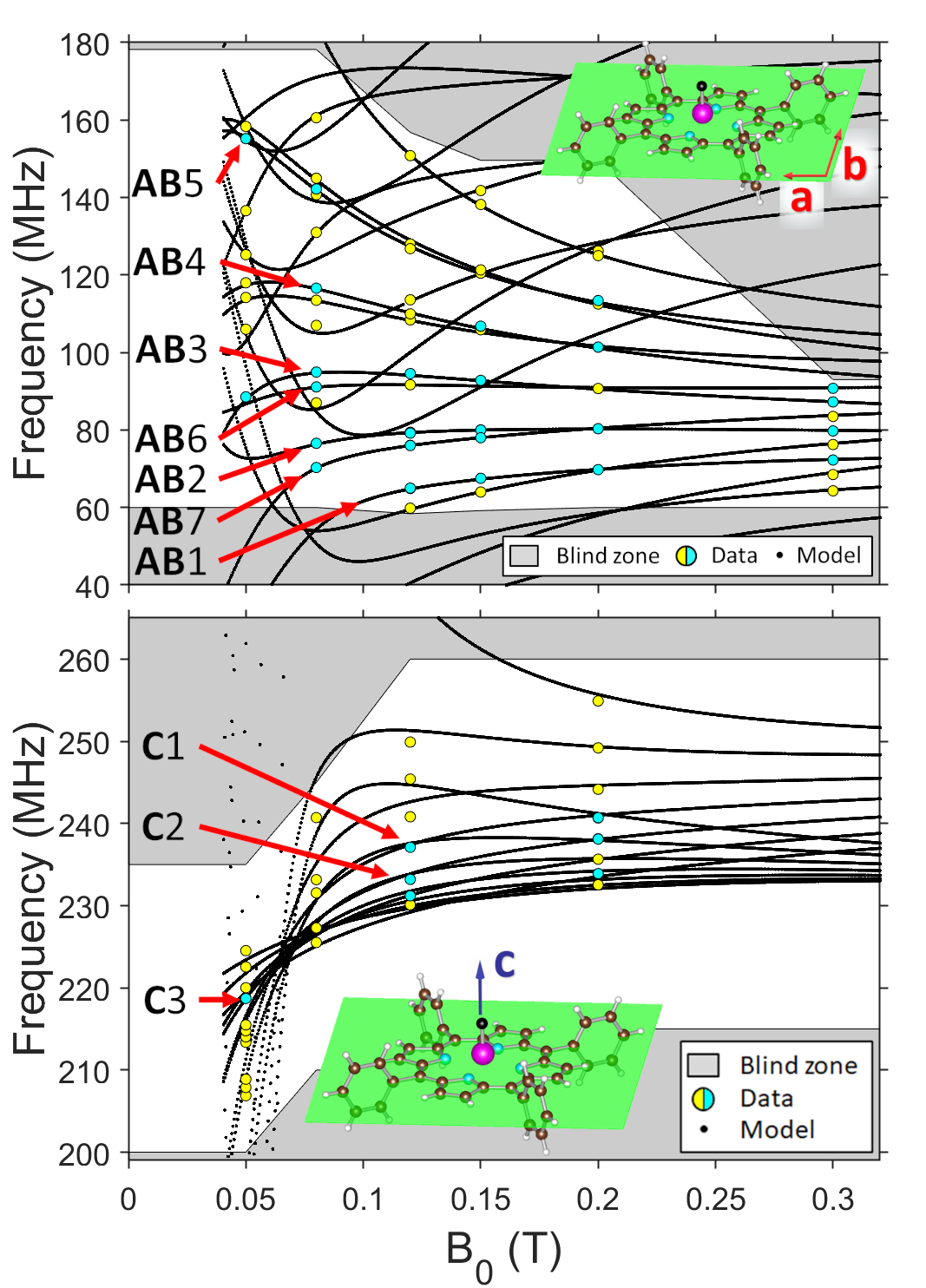} \\
\caption{Measured (yellow/cyan dots, $T=1.4$ K) and calculated (black lines) transition frequencies as a function of the applied field, applied along the directions depicted in the inset: $B_0 \in \emph{ab}$-plane and $B_0 \parallel \boldsymbol{\hat{c}}$, respectively.  Experimental data represented by cyan dots and labelled $"\boldsymbol{AB} \#"$ represent peaks for which the $T_2$ nuclear spin dephasing time is measured. Shaded gray areas are experimentally non-accessible or not explored.}
\label{Fig1}
\end{figure}
\begin{figure}[t]
\centering
\includegraphics[width=1\linewidth]{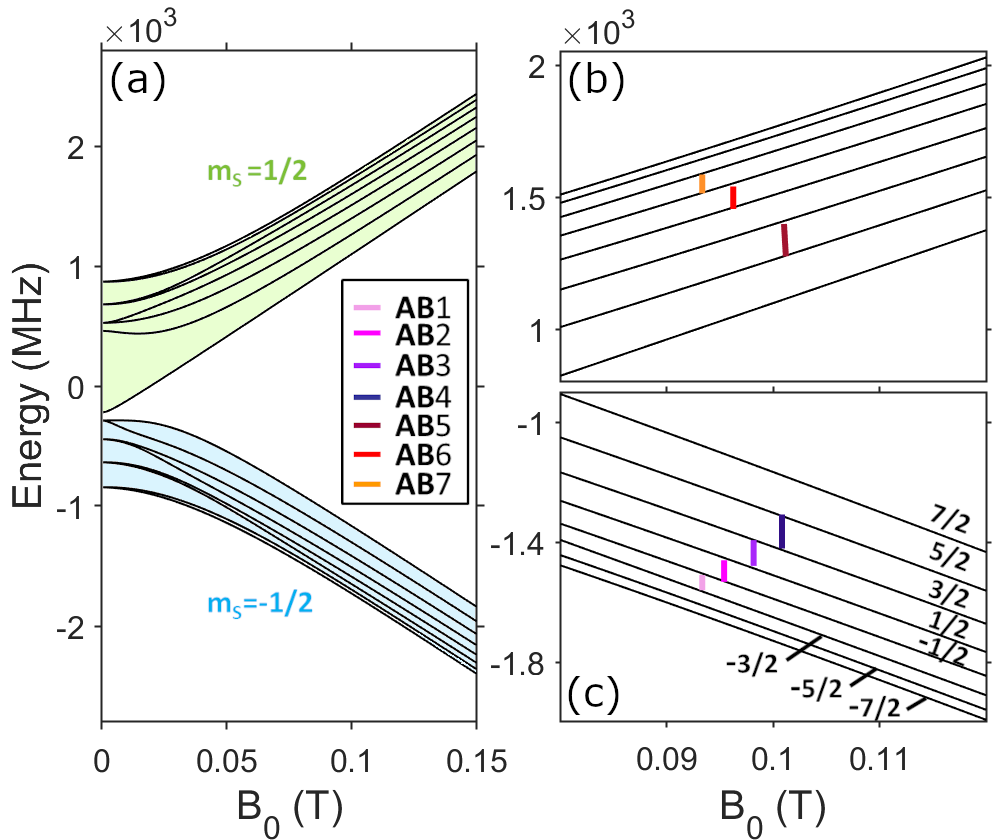}
\caption{Energy level diagram as a function of the static field, applied in the \emph{ab} plane. In the left panel (a), green and light blue shaded areas highlight the $m_{S}=1/2$ and -1/2 spin multiplets, respectively, while in the right panels (b,c), the nuclear levels of the lowest electronic spin manifold are labelled by the corresponding nuclear spin component along the field $m_{I}$. Vertical coloured marks indicate nuclear transitions for which the nuclear spin-spin relaxation time $T_2$ is measured.}
\label{Levels}
\end{figure}
The system is described by the spin Hamiltonian:
\begin{equation} \label{Ham}
H_0 = \mathbf{\hat{I}\cdot A\cdot\hat{S}} + p \hat{I}_z^2+\mu_B \mathbf{\hat{S}} \cdot \mathbf{g}_S \cdot \mathbf{B}_0 +\mu_N g_N \mathbf{\hat{I}} \cdot \mathbf{ B}_0
\end{equation}
where the first term represents the hyperfine interaction between the nuclear spin 7/2 and the electronic spin 1/2, the second is the nuclear quadrupolar interaction, and the last two terms model the electronic and nuclear Zeeman terms. 
In the orthogonal reference frame defined by the crystallographic axes 
($abc \equiv xyz$) 
all tensors are diagonal, collinear, and show axial symmetry.
The electronic $\mathbf{g}_S$ tensor was already characterized in previous EPR studies \cite{VOTPP}, with $g_{{x,y}} = 1.9865,\ g_{z} = 1.963$ and was thus kept fixed in the following refinement. Analogously, we fix $\mu_N g_N$ to the known value of $-11.213$ MHz/T for $^{51}$V \cite{TableG}.  \\
Conversely, hyperfine tensor ({\bf A}) and quadrupolar interaction term ($p$) are fitted to reach the best agreement between simulated and measured spectra, 
as a function of the applied magnetic field (figure \ref{Fig1}). 
We find $A_{x,y}=-170\pm 1$ MHz, $A_{z}=-480\pm 1$ MHz, $p=-0.35\pm 0.07$ MHz. \\
The full set of observed NMR peaks as a function of the static field $B_0$ is reported in Figure \ref{Fig1}, in both the measured directions. We note that experimental transition frequencies are in very good agreement with simulations using spin Hamiltonian (\ref{Ham}) and best fit parameters listed above. \\ 
The resulting energy levels diagram, obtained from diagonalization of spin Hamiltonian (\ref{Ham}) with magnetic field applied in the \emph{ab} plane, is shown in figure \ref{Levels}. Despite the mixing between electronic and nuclear spins induced by the significant transverse components of the hyperfine interaction ($A_{y,z}$) with respect to the static field direction, for $B_0 \gtrsim 0.25$ T,  $g_x\mu_BB_0>|A_z|$ and the eigenstates are $98\%$ factorized for $B_0=0.25$ T. In these conditions, the eigenstates can be labeled by the components of $S$ and $I$ parallel to the external field,  $\ket{m_{S},m_{I}}$. \\ 

\subsection{Phase memory time}
\label{section:T2&Rabi}
\begin{figure}[t]
\centering
  \includegraphics[width=1\linewidth]{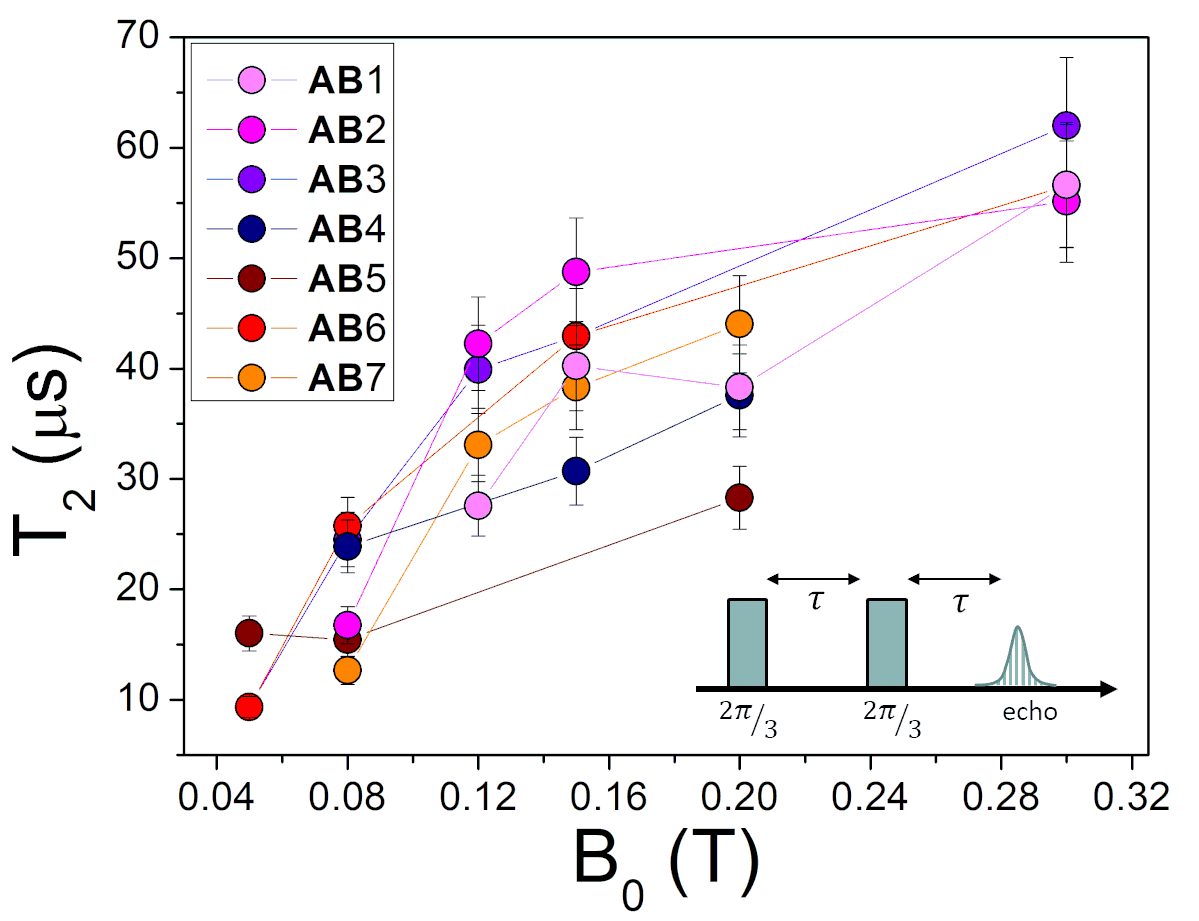}
\caption{Nuclear spin dephasing times $T_2^{m_I,m_I^\prime}$, measured at $T=1.4$ K, with the Han-echo sequence depicted in the inset, as a function of the applied field $B_0$ in the \emph{ab} plane, for transition highlighted in figure \ref{Levels}-(b,c).}
\label{T2}
\end{figure}

We now focus on the experimental configuration with the field lying in the \emph{ab} plane, which is the one ensuring the best resolution between nuclear spin transitions (see discussion below).
We investigate nuclear spin dephasing rates $1/T_2^{m_I,m_I^\prime}$ for different $m_I \rightarrow m_I^\prime$ transitions (vertical colored lines in Fig. \ref{Levels}-(b,c)), as a function of the applied field. Analogous results with $B_0$ parallel to $c$ are reported in the SI (Figure S4).
The phase memory time is extracted from the decay of the spin-echo amplitude $M(\tau)$
after application of the Hahn echo sequence $2\pi/3-\tau-2\pi/3$ (see SI), 
as a function of the delay $\tau$ between the exciting and refocusing pulses \cite{Han}. \\
These decays are accurately fitted by a single exponential function (Figure S4).
The corresponding phase memory times $T_2^{m_I,m_I^\prime}$ are reported in Figure \ref{T2}, as a function of the applied field. 
We find $T_2^{m_I,m_I^\prime}$ increasing from 10 to 60 $\mu s$ with increasing $B_0$, for all transitions. \\
Since the dilution of [VO(TPP)] in the diamagnetic [TiO(TPP)] matrix strongly suppresses the electron-electron spin dipolar coupling, the decoherence of the central $^{51}$V nuclear spin is primarily ruled by the interaction with neighbouring nuclei.
For the low magnetic fields examined here, this interaction is mediated by  virtual excitations of the electronic spin component of the system wave-function, 
whereas direct coupling between nuclear spins is much smaller. As a consequence, the reduction of electron-nuclear mixing with increasing $B_0$ is responsible of the observed increase of $T_2$ with magnetic field.
A secondary source of decoherence (limited by dilution) is represented by the interaction of the probed nuclei with surrounding electronic spins. This effect is also reduced by increasing $B_0$, due to the consequent increment in electronic polarization and suppression of  electronic spins fluctuations. \\

\subsection{Coherent manipulation by rf pulses}
Having demonstrated a remarkable coherence for several nuclear spin transitions, we now show that the investigated system also fulfills the second requirement to build a good qudit: individual addressability of these transitions. 
Indeed, we show below the capability to coherently manipulate the system, demonstrating that selective excitation of each of these transitions by resonant rf pulses drive Rabi oscillations between nuclear levels, conditioned by the electronic spin state.
We focus, in particular, on the $m_S=-1/2$ manifold, which constitutes the relevant computational subspace for algorithms exploiting the electronic spin as an ancilla of the nuclear qudit, see e.g. Ref. \cite{JPCLqec}. 
To individually address each nuclear transition, the difference between nuclear gaps $\delta(m_I) = (E_{m_I+1}-E_{m_I}) - (E_{m_I}-E_{m_I-1})$ must be significantly larger than the spectral width of the excitation pulses.
If we only retain secular terms in the spin Hamiltonian (i.e. for large $B_0$ or small transverse hyperfine couplings), such a difference is due to nuclear quadrupolar interaction ($\delta_{m_I}=2p$ for $B_0$ parallel to $z$), which however is very small in the examined $^{51}$V system.
Small values of $p$ occur also in other 3$d$-ion based molecular nanomagnets. Here we demonstrate that those small quadrupolar terms are not a limiting factor in terms of addressability of nuclear transitions. Indeed, by properly tuning the applied field, we can meet a condition in which the significant transverse hyperfine interaction grants a sizeable second-order pseudo-quadrupolar contribution to the system at low applied field (see SI).
\begin{figure}[ht!]
\centering
\includegraphics[width=1\linewidth]{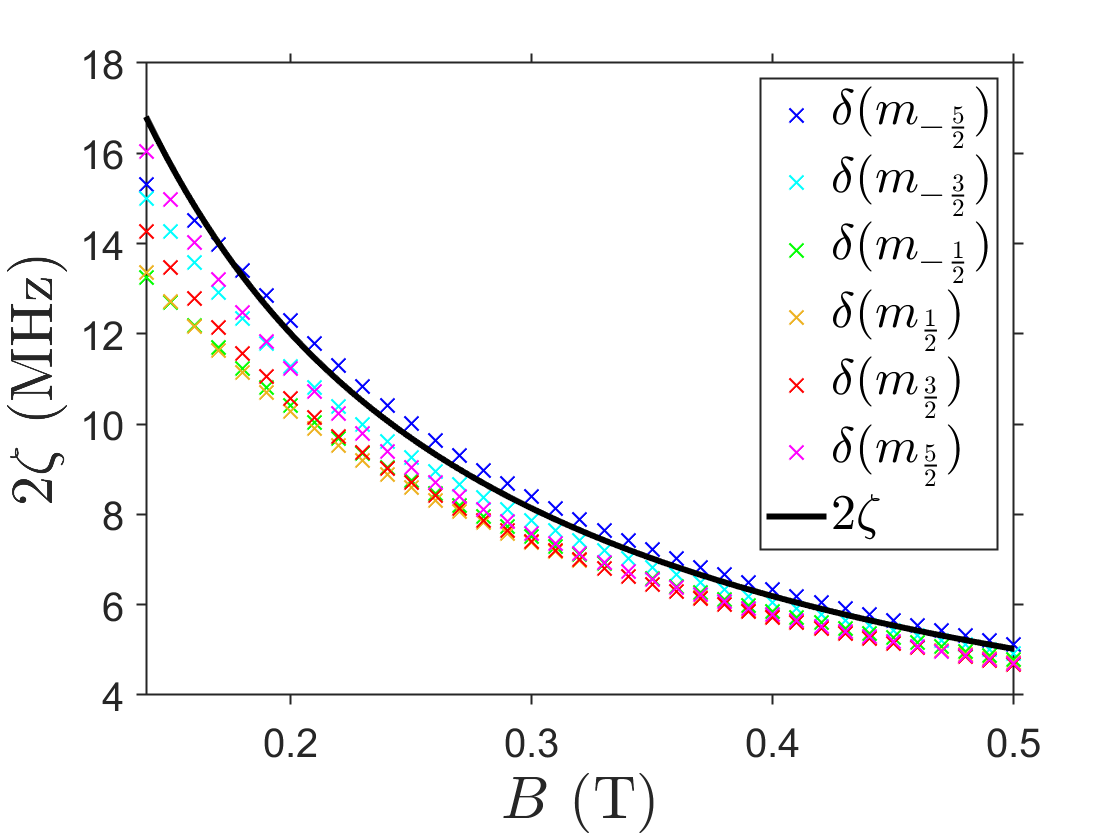}
\caption{Difference between the energy gap of consecutive nuclear transitions $\delta(m_I) = (E_{m_I+1}-E_{m_I}) - (E_{m_I}-E_{m_I-1})$ in the $m_S=-1/2$ subspace,
as a function of static field, applied along $x$. Results from diagonalization of the effective rank-2 pseudo-quadrupolar Hamiltonian (\ref{Hq}) are shown in black, in very good agreement with results from exact diagonalization of the full Hamiltonian (\ref{Ham}) (colored crosses), especially at large $B_0$.}
\label{Peff}
\end{figure}
This effective quadrupolar coupling is obtained by considering the effect of hyperfine interaction up to second-order perturbation theory with respect to the Zeeman term due to a static field $B_0$, and is maximized by applying the external field along $x$. Indeed, with this choice the components of {\bf A} perpendicular to {\bf B}$_0$ are the largest.
The resulting low-energy Hamiltonian reads (apart from a constant)
\begin{equation}
    H_{q} = p_{\rm eff}^x [I_x^2 - I(I+1)/3] + p_{\rm eff}^r (I_z^2-I_y^2) + B_{\rm eff} I_x
    \label{Hq}
\end{equation}
where
\begin{eqnarray} \nonumber
    p_{\rm eff}^x &=& \frac{ (A_z^2+A_y^2) g_e \mu_B + 2 A_y A_z g_N \mu_N}{8 B_0( g_e^2 \mu_B^2 - g_N^2 \mu_N^2)}  \\
    p_{\rm eff}^r &=& \frac{ (A_y^2-A_z^2) g_e \mu_B}{8 B_0( g_e^2 \mu_B^2 - g_N^2 \mu_N^2)} \\ \nonumber
    B_{\rm eff} &=& \frac{A_x}{2} +  \frac{ (A_z^2+A_y^2) g_N \mu_N + 2 A_y A_z g_e \mu_B}{8 B_0( g_e^2 \mu_B^2 - g_N^2 \mu_N^2)}
\end{eqnarray}
The last term is an effective magnetic field with both first and second-order perturbation theory contributions. 
For sake of simplicity, we have neglected in this analysis the intrinsic quadrupole coupling, which is found to be significantly smaller than $p_{\rm eff}$ in the examined $B_0$ range.
The average difference ($2\zeta$) between the energy gaps obtained from 
diagonalization of $H_q$ represents an effective quadrupole interaction which allows us to resolve different transitions and is larger at low $B_0$.
It is reported in Fig. \ref{Peff}, compared to the result obtained from diagonalization of the full Hamiltonian (\ref{Ham}) for the different gaps ($\delta(m_I)$). 
The agreement is good and it improves at larger fields, where  
the electronic-nuclear mixing of spin eigenstates is reduced and the perturbative expansion becomes more accurate. 
\begin{figure} [t]
\centering
\includegraphics[width=1\linewidth]{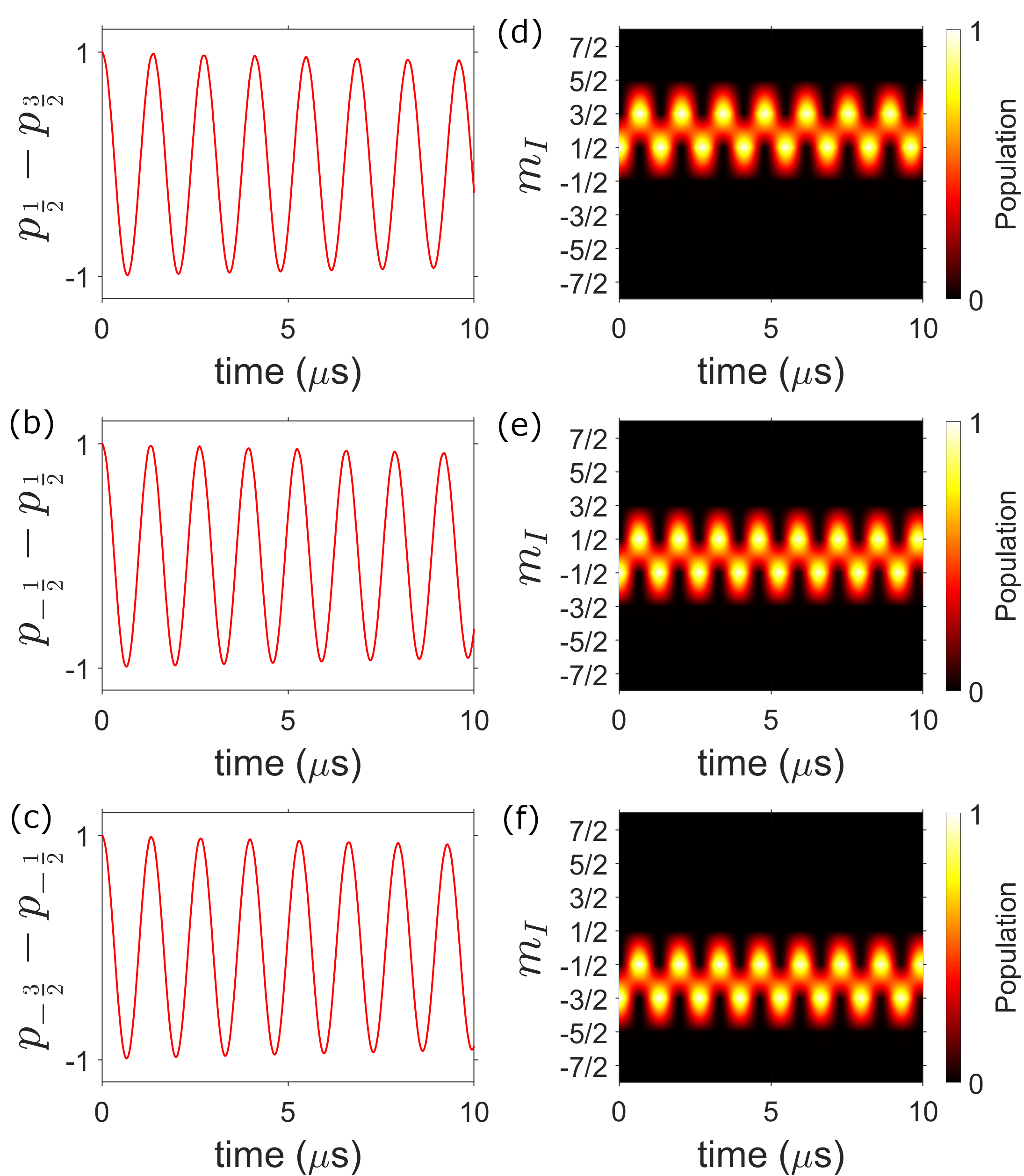}
\caption{Simulated time evolution of the population of the different nuclear spin levels, within $m_S=-1/2$ subspace, i.e. $p_{m_I}=\langle m_I,m_S=-1/2| \, \rho \, | m_I,m_S=-1/2 \rangle$. The static field is applied along $x$ and quantum numbers $m_I,m_S$ are eigenvalues of $I_x,S_x$.
The system is prepared in specific initial states and addressed by an oscillating field resonant with specific $\Delta m_I = 1$ transitions: (a,d) $m_I=-3/2\rightarrow -1/2$, (b,e) $m_I=-1/2\rightarrow 1/2$, (c,f) $m_I=1/2\rightarrow 3/2$.
Left panels (a-c): difference between population of the two addressed levels.
Right (d-f): 2D heat-maps of the population fraction for each nuclear level, showing that  only the targeted levels exhibit sizable oscillations, while  leakage to other levels is negligible $\sim 0$.
Several Rabi oscillations are shown, including the effect of pure dephasing with the measured values of $T_2^{m_I,m_I^\prime}$.}
\label{Fig4}
\end{figure}
\begin{figure}[ht!]
\centering
\includegraphics[width=1\linewidth]{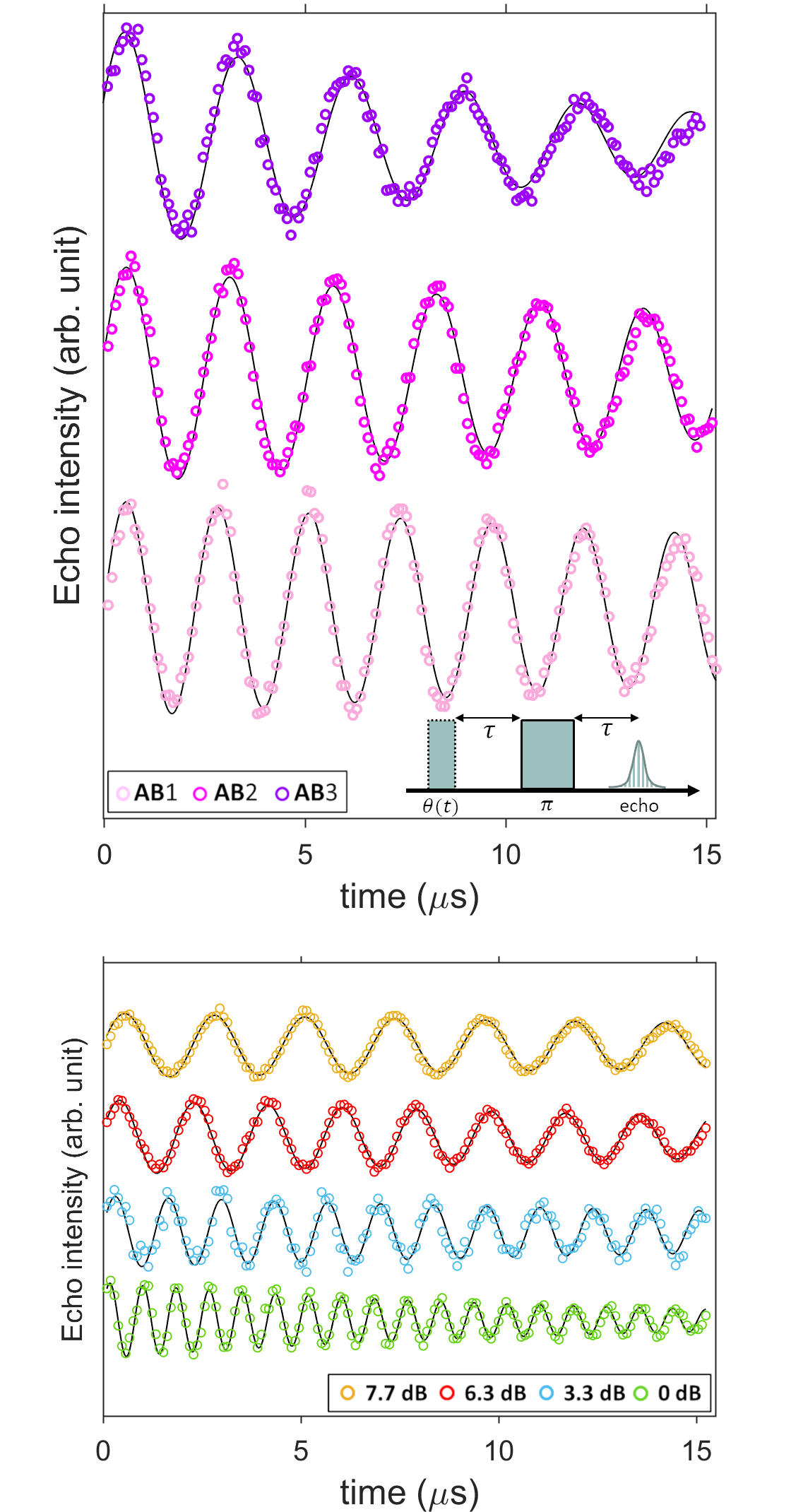}
\caption{(a) Nuclear Rabi oscillations induced on transitions $\boldsymbol{AB} \ 1,2,3$ at fixed static field $B_0=0.3$ T,
approximately parallel to $x$ ($\sim 2^\circ$ uncertainty)
by the linearly polarized ($z$) rf pulse sequence depicted in the inset (with variable first pulse length $\theta(t)$) and fixed radio frequency attenuation 7.7 dB. (b) Nuclear Rabi oscillations induced on transition $\boldsymbol{AB} 1$ ($f=72.25$ MHz) at fixed field ($B_0=0.3$ T) for different applied pulse intensities, showing the scaling of Rabi frequency with $B_1$ (figure S5-S6) Experiments are performed at $T=1.4$ K.}
\label{Fig5}
\end{figure}
To demonstrate that we can actually individually address the $\Delta m_I = \pm 1$ nuclear spin transitions, thus inducing monochromatic oscillations between specific pairs of levels, we simulate the time evolution of the system subject to a rf pulse resonant with the desired transition. In particular, we start from the Hamiltonian extracted in section \ref{section:HamiltonianFitting}, whose parameters are sound given the very good agreement between calculated and measured spectra, and we numerically solve the Lindblad equation for the system density matrix $\rho$:
\begin{align}
\nonumber
 \frac{d\rho}{dt} & = -i[H_0+H_1(t),\rho]  \\  \nonumber
& + \sum_{m_I,m_I^\prime} \gamma_{m_I} \delta_{m_I,m_I^\prime} [\, |m_I \rangle \langle m_I | \rho | m_I \rangle \langle m_I | \\
& -  (|m_I \rangle \langle m_I| \rho + \rho |m_I \rangle \langle m_I|)/2 \, ]
\end{align}
Here the first term represents the coherent evolution of the system according to the static Hamiltonian $H_0$ and the excitation pulse $H_1(t)$, while the second models nuclear spin dephasing mechanism, with dephasing rates $1/T_2^{m_I m_I^\prime} = (\gamma_{m_I}+\gamma_{m_I^\prime} )/2$ for each nuclear transition $m_I \rightarrow m_I^\prime$.
Our simulations employ the Hamiltonian parameters fitted in \ref{section:HamiltonianFitting} and the measured dephasing times $1/T_2^{m_I m_I^\prime}$, thus perfectly matching the experimental scenario.\\
As an example, we consider transitions between nuclear spin levels $m_I=-\frac{3}{2},-\frac{1}{2},\frac{1}{2},\frac{3}{2}$, belonging to the $m_S = -\frac{1}{2}$ multiplet (labelled \textbf{AB}1,2,3 in the energy levels diagram of figure \ref{Levels}).
These four levels, combined with the electronic spin doublet, could be used to implement the quantum error correction protocol proposed in Ref. \cite{JPCLqec}.\\
Simulations (Fig \ref{Fig4}) are performed with $B_0 = 0.3$ T $\hat{{\bf x}}$, which constitutes a good trade-off between electro-nuclear factorization and spectral separation of the different transitions. 
The simulation shows that by using linearly polarized rectangular pulses of amplitude ${\bf B}_1 = 5$ G $\hat{{\bf z}}$ the system prepared in state $m_I=-3/2$ (Fig. \ref{Fig4}-c) evolves correctly into $m_I = -1/2$, with negligible leakage to neighboring levels. An analogous performance is found for transitions from $m_I=-1/2$ to $m_I=1/2$ [Fig. \ref{Fig4}-(b)] and from $m_I=1/2$ to $m_I=3/2$ [Fig. \ref{Fig4}-(a)]. 
In order to demonstrate the ability to selectively excite each transition in the experimental conditions we finally simulate Rabi experiments
using an initial state with population on all nuclear states belonging to the $m_S=-1/2$ manifold. We find that for $B_1 \approx 5$ G, only the targeted levels populations are significantly modified (see SI and Figure S7). 
This value of $B_1$ allows us to
obtain monochromatic Rabi oscillations between each pair of addressed $m_I$ levels (i.e. with negligible leakage to other states) with the shortest possible period, thus reducing the effect of decoherence. 
In addition, with the chosen ${\bf B}_0$ the electron-nuclear mixing grants large matrix elements of the addressed transitions, while still keeping factorization of the system wave-function. Hence, in these conditions both the effect of decoherence (which yields a damping of the oscillations reported in Fig. \ref{Fig4}) and the unwanted leakage to other states are very small (see panels (d-f)). \\
The simulated excitation pulse represents a feasible experimental condition, and can be easily reproduced using our NMR spectrometer (see SI). Thus, to validate the results of the simulations, we target the same transitions in a pulse NMR experiment consisting of a first
exciting pulse $P_{\theta}$ of increasing length, followed by a fixed $P_{\pi}$ refocusing pulse (details are reported in the SI).  
To ensure coherent selective manipulation of just the two addressed levels, we have performed nutation experiments with weak pulses, 
as shown in figure \ref{Fig5}. The resulting oscillations are modelled by an exponentially damped sinusoidal function $f(t)\propto e^{-t \lambda} {\rm sin} (2\pi \nu_R t)$.
For each excited transition, the measured Rabi frequency $\nu_R$ increases linearly with oscillating field amplitude $B_1$ in agreement with expectations for an rf field induced nutation (see SI). \\
The damping rate ($\lambda$) of the measured Rabi oscillations shows a strong dependence on $B_1$, indicating that this decay is dominated by inhomogeneities of the applied radiofrequency field. However, for achievable pulse intensities the decay time $\frac{1}{\lambda}$ is always remarkably longer that the duration of a $\pi$ rotation, in the range $0.3-1 \mu$ s, as shown in Fig. \ref{Fig5} and also in Fig. S5.
The long decay times allow many successive operations of nuclear spin state manipulation  before the coherence is lost. \\


\section{Conclusions}
	
Summarizing, we have shown that	broadband nuclear magnetic resonance allows us to extract a very accurate description of the coupled $^{51}$V nuclear spin 7/2 and electron spin 1/2 levels in the [VO(TPP)] molecular complex, and to coherently manipulate it.  \\
The system represents an 8-level nuclear qudit coupled to an electronic qubit, particularly promising for quantum computing applications. Indeed, quantum simulators and quantum error-correction schemes based on such qubit-qudit elementary units have been recently proposed \cite{JPCLqec}, and several other algorithms could benefit from the larger number of available levels offered by a qudit, compared to its spin 1/2 counterpart \cite{RevQudits}. To achieve this, we need long coherence times and individual addressability for each nuclear spin transition.  \\
We demonstrate that these conditions are fulfilled by the investigated system by an extensive NMR study on single crystals. In particular, we first characterize the nuclear spin Hamiltonian by fitting NMR spectra in a wide frequency range, for different orientations and intensities of the applied magnetic field. In addition, we probe nuclear coherences by Hahn echo experiments to extract the dephasing rates of nuclear excitations as a function of the applied field. 
This gives us a solid starting point for simulating the dynamics of the system and for designing pulse NMR experiments to manipulate the different nuclear transitions. 
Indeed, by combining numerical simulations to targeted measurements, we have shown that for properly chosen experimental conditions, individual nuclear transitions can be adressed with neiglible leackage vs other transitions.
This is made possible by the remarkably long coherence times and by the robust hyperfine interaction, which induces an effective quadrupolar coupling distinguishing the various transitions and hence enables 
and coherent manipulation of the qudit. \\
Such an effective quadrupolar coupling could significantly enlarge the list of suitable qubit-qudit systems for quantum computing applications. In these conditions, not only rare-earth \cite{jacsYb}, but also transition metal complexes could be very promising, thanks to their remarkable coherence times, despite their smaller intrinsic quadrupole interaction. 
In particular, V-based compounds, usually not considered due to their small quadrupole coupling, become very attractive, by combining two great advantages. First,
among all transition-metal and rare-earth MNMs, they possess 
the largest nuclear spin (7/2), thus providing a sizable number of levels for the implementation of qudit algorithms.
Second, the record electronic coherence for a MNM was reported on a V complex \cite{Zadrozny}, which is only an example of a large family of highly coherent compounds \cite{Freedman_JACS,Atzori2016,Phonons}.  
The [VO(TPP)] case examined here shows indeed values of nuclear $T_2$ significantly longer than those reported in Ref. \cite{jacsYb} for [Yb(trensal]), whose nuclear spin was characterized by a much stronger value of $p$.\\
It is finally worth noting that the examined compound presents some additional appealing features. [VO(TPP)], though seldom investigated, belongs to a class of molecules that can be evaporated and organized in ordered arrays on different substrates. \cite{SurfaceGOTTFRIED} The V=O group is readily accessible  in these flat molecules \cite{FlatMol}, with the potential to perform single spin EPR by using a Scanning Probe Microscopy, as recently reported for a phthalocyanine complex \cite{EPRScanningProbe}.
Equally relevant is the possibility to couple more porphyrin rings to induce interactions between the coordinated metal ions \cite{IonsInteract},  as required for the implementation of quantum gates. Last but not least, chemical functionalization of the porphyrin allowed to arrange the [VO(TPP)] units into a metal-organic framework without affecting its magnetic properties \cite{VOTPP}.
This, together with its remarkable coherence times, makes it a good candidate for a magnetic field sensor. To this aim, the hyperfine coupling with a spin 7/2 nucleus could be a useful resource to enhance sensitivity of a [VO(TPP)]-based quantum meter close to level anti-crossings  \cite{PRBTroiani2016,PRAGhirardi2018,RMPSensing}, such as those occurring below 0.05 T. \\ 
	
\section*{Acknowledgements}
This work has received funding from the European Union’s Horizon 2020 research and innovation programme (FET-OPEN project FATMOLS) under grant agreement No 862893, the European Project ``Scaling Up quantum computation with MOlecular spins" (SUMO) of the call QuantERA, cofunded by Italian Ministry of Education and Research (MUR), and by the Italian Ministry of Education and Research (MUR) through PRIN Project 2017 Q-chiSS  ``Quantum detection of chiral-induced spin selectivity at the molecular level''. \\


\begin{thebibliography}{62}%
	\makeatletter
	\providecommand \@ifxundefined [1]{%
		\@ifx{#1\undefined}
	}%
	\providecommand \@ifnum [1]{%
		\ifnum #1\expandafter \@firstoftwo
		\else \expandafter \@secondoftwo
		\fi
	}%
	\providecommand \@ifx [1]{%
		\ifx #1\expandafter \@firstoftwo
		\else \expandafter \@secondoftwo
		\fi
	}%
	\providecommand \natexlab [1]{#1}%
	\providecommand \enquote  [1]{``#1''}%
	\providecommand \bibnamefont  [1]{#1}%
	\providecommand \bibfnamefont [1]{#1}%
	\providecommand \citenamefont [1]{#1}%
	\providecommand \href@noop [0]{\@secondoftwo}%
	\providecommand \href [0]{\begingroup \@sanitize@url \@href}%
	\providecommand \@href[1]{\@@startlink{#1}\@@href}%
	\providecommand \@@href[1]{\endgroup#1\@@endlink}%
	\providecommand \@sanitize@url [0]{\catcode `\\12\catcode `\$12\catcode
		`\&12\catcode `\#12\catcode `\^12\catcode `\_12\catcode `\%12\relax}%
	\providecommand \@@startlink[1]{}%
	\providecommand \@@endlink[0]{}%
	\providecommand \url  [0]{\begingroup\@sanitize@url \@url }%
	\providecommand \@url [1]{\endgroup\@href {#1}{\urlprefix }}%
	\providecommand \urlprefix  [0]{URL }%
	\providecommand \Eprint [0]{\href }%
	\providecommand \doibase [0]{https://doi.org/}%
	\providecommand \selectlanguage [0]{\@gobble}%
	\providecommand \bibinfo  [0]{\@secondoftwo}%
	\providecommand \bibfield  [0]{\@secondoftwo}%
	\providecommand \translation [1]{[#1]}%
	\providecommand \BibitemOpen [0]{}%
	\providecommand \bibitemStop [0]{}%
	\providecommand \bibitemNoStop [0]{.\EOS\space}%
	\providecommand \EOS [0]{\spacefactor3000\relax}%
	\providecommand \BibitemShut  [1]{\csname bibitem#1\endcsname}%
	\let\auto@bib@innerbib\@empty
	\bibitem [{\citenamefont {Arute}\ and\ \citenamefont
		{et~al.}(2019)}]{Supremacy}%
	\BibitemOpen
	\bibfield  {author} {\bibinfo {author} {\bibfnamefont {F.}~\bibnamefont
			{Arute}}\ and\ \bibinfo {author} {\bibnamefont {et~al.}},\ }\bibfield
	{title} {\bibinfo {title} {Quantum supremacy using a programmable
			superconducting processor.},\ }\href
	{https://doi.org/https://doi.org/10.1038/s41586-019-1666-5} {\bibfield
		{journal} {\bibinfo  {journal} {Nature}\ }\textbf {\bibinfo {volume} {574}},\
		\bibinfo {pages} {505} (\bibinfo {year} {2019})}\BibitemShut {NoStop}%
	\bibitem [{\citenamefont {Tacchino}\ \emph {et~al.}(2019)\citenamefont
		{Tacchino}, \citenamefont {Chiesa}, \citenamefont {Carretta},\ and\
		\citenamefont {Gerace}}]{RevTacchino}%
	\BibitemOpen
	\bibfield  {author} {\bibinfo {author} {\bibfnamefont {F.}~\bibnamefont
			{Tacchino}}, \bibinfo {author} {\bibfnamefont {A.}~\bibnamefont {Chiesa}},
		\bibinfo {author} {\bibfnamefont {S.}~\bibnamefont {Carretta}},\ and\
		\bibinfo {author} {\bibfnamefont {D.}~\bibnamefont {Gerace}},\ }\bibfield
	{title} {\bibinfo {title} {Quantum computers as universal quantum simulators:
			state-of-art and perspectives.},\ }\href
	{https://doi.org/https://doi.org/10.1002/qute.201900052} {\bibfield
		{journal} {\bibinfo  {journal} {Adv. Quantum Technol.}\ ,\ \bibinfo {pages}
			{1900052}} (\bibinfo {year} {2019})}\BibitemShut {NoStop}%
	\bibitem [{\citenamefont {Chiesa}\ \emph {et~al.}(2019)\citenamefont {Chiesa},
		\citenamefont {Tacchino}, \citenamefont {Grossi}, \citenamefont {Santini},
		\citenamefont {Tavernelli}, \citenamefont {Gerace},\ and\ \citenamefont
		{Carretta}}]{NatPhysIBM}%
	\BibitemOpen
	\bibfield  {author} {\bibinfo {author} {\bibfnamefont {A.}~\bibnamefont
			{Chiesa}}, \bibinfo {author} {\bibfnamefont {F.}~\bibnamefont {Tacchino}},
		\bibinfo {author} {\bibfnamefont {M.}~\bibnamefont {Grossi}}, \bibinfo
		{author} {\bibfnamefont {P.}~\bibnamefont {Santini}}, \bibinfo {author}
		{\bibfnamefont {I.}~\bibnamefont {Tavernelli}}, \bibinfo {author}
		{\bibfnamefont {D.}~\bibnamefont {Gerace}},\ and\ \bibinfo {author}
		{\bibfnamefont {S.}~\bibnamefont {Carretta}},\ }\bibfield  {title} {\bibinfo
		{title} {Quantum hardware simulating four-dimensional inelastic neutron
			scattering.},\ }\href
	{https://doi.org/https://doi.org/10.1038/s41567-019-0437-4} {\bibfield
		{journal} {\bibinfo  {journal} {Nature Phys.}\ }\textbf {\bibinfo {volume}
			{15}},\ \bibinfo {pages} {455} (\bibinfo {year} {2019})}\BibitemShut
	{NoStop}%
	\bibitem [{\citenamefont {Arute}\ and\ \citenamefont
		{et~al.}(2020)}]{HFGoogle}%
	\BibitemOpen
	\bibfield  {author} {\bibinfo {author} {\bibfnamefont {F.}~\bibnamefont
			{Arute}}\ and\ \bibinfo {author} {\bibnamefont {et~al.}},\ }\bibfield
	{title} {\bibinfo {title} {Hartree-fock on a superconducting qubit quantum
			computer.},\ }\href {https://doi.org/10.1126/science.abb9811} {\bibfield
		{journal} {\bibinfo  {journal} {Science}\ }\textbf {\bibinfo {volume}
			{369}},\ \bibinfo {pages} {1084} (\bibinfo {year} {2020})}\BibitemShut
	{NoStop}%
	\bibitem [{\citenamefont {Ollitrault}\ \emph {et~al.}(2020)\citenamefont
		{Ollitrault}, \citenamefont {Baiardi}, \citenamefont {Reiher},\ and\
		\citenamefont {Tavernelli}}]{D0SC01908A}%
	\BibitemOpen
	\bibfield  {author} {\bibinfo {author} {\bibfnamefont {P.~J.}\ \bibnamefont
			{Ollitrault}}, \bibinfo {author} {\bibfnamefont {A.}~\bibnamefont {Baiardi}},
		\bibinfo {author} {\bibfnamefont {M.}~\bibnamefont {Reiher}},\ and\ \bibinfo
		{author} {\bibfnamefont {I.}~\bibnamefont {Tavernelli}},\ }\bibfield  {title}
	{\bibinfo {title} {Hardware efficient quantum algorithms for vibrational
			structure calculations},\ }\href {https://doi.org/10.1039/D0SC01908A}
	{\bibfield  {journal} {\bibinfo  {journal} {Chem. Sci.}\ }\textbf {\bibinfo
			{volume} {11}},\ \bibinfo {pages} {6842} (\bibinfo {year}
		{2020})}\BibitemShut {NoStop}%
	\bibitem [{\citenamefont {Kandala}\ \emph {et~al.}(2017)\citenamefont
		{Kandala}, \citenamefont {Mezzacapo}, \citenamefont {Temme}, \citenamefont
		{Takita}, \citenamefont {Brink}, \citenamefont {Chow},\ and\ \citenamefont
		{Gambetta}}]{vqe1}%
	\BibitemOpen
	\bibfield  {author} {\bibinfo {author} {\bibfnamefont {A.}~\bibnamefont
			{Kandala}}, \bibinfo {author} {\bibfnamefont {A.}~\bibnamefont {Mezzacapo}},
		\bibinfo {author} {\bibfnamefont {K.}~\bibnamefont {Temme}}, \bibinfo
		{author} {\bibfnamefont {M.}~\bibnamefont {Takita}}, \bibinfo {author}
		{\bibfnamefont {M.}~\bibnamefont {Brink}}, \bibinfo {author} {\bibfnamefont
			{J.~M.}\ \bibnamefont {Chow}},\ and\ \bibinfo {author} {\bibfnamefont
			{J.~M.}\ \bibnamefont {Gambetta}},\ }\bibfield  {title} {\bibinfo {title}
		{Hardware-efficient variational quantum eigensolver for small molecules and
			quantum magnets},\ }\href {https://doi.org/10.1038/nature23879} {\bibfield
		{journal} {\bibinfo  {journal} {Nature}\ }\textbf {\bibinfo {volume} {549}},\
		\bibinfo {pages} {242–246} (\bibinfo {year} {2017})}\BibitemShut {NoStop}%
	\bibitem [{\citenamefont {Bruzewicz}\ \emph {et~al.}(2019)\citenamefont
		{Bruzewicz}, \citenamefont {Chiaverini}, \citenamefont {McConnell},\ and\
		\citenamefont {Sage}}]{RevTrappedIons}%
	\BibitemOpen
	\bibfield  {author} {\bibinfo {author} {\bibfnamefont {C.~D.}\ \bibnamefont
			{Bruzewicz}}, \bibinfo {author} {\bibfnamefont {J.}~\bibnamefont
			{Chiaverini}}, \bibinfo {author} {\bibfnamefont {R.}~\bibnamefont
			{McConnell}},\ and\ \bibinfo {author} {\bibfnamefont {J.~M.}\ \bibnamefont
			{Sage}},\ }\bibfield  {title} {\bibinfo {title} {Trapped-ion quantum
			computing: Progress and challenges},\ }\href@noop {} {\bibfield  {journal}
		{\bibinfo  {journal} {Appl. Phys. Lett.}\ }\textbf {\bibinfo {volume} {6}},\
		\bibinfo {pages} {021314} (\bibinfo {year} {2019})}\BibitemShut {NoStop}%
	\bibitem [{\citenamefont {Preskill}(2018)}]{NISQ}%
	\BibitemOpen
	\bibfield  {author} {\bibinfo {author} {\bibfnamefont {J.}~\bibnamefont
			{Preskill}},\ }\bibfield  {title} {\bibinfo {title} {Quantum computing in the
			nisq era and beyond.},\ }\href
	{https://doi.org/https://doi.org/10.22331/q-2018-08-06-79} {\bibfield
		{journal} {\bibinfo  {journal} {Quantum}\ }\textbf {\bibinfo {volume} {2}},\
		\bibinfo {pages} {79} (\bibinfo {year} {2018})}\BibitemShut {NoStop}%
	\bibitem [{\citenamefont {Wang}\ \emph {et~al.}(2020)\citenamefont {Wang},
		\citenamefont {Hu}, \citenamefont {Sanders},\ and\ \citenamefont
		{Kais}}]{RevQudits}%
	\BibitemOpen
	\bibfield  {author} {\bibinfo {author} {\bibfnamefont {Y.}~\bibnamefont
			{Wang}}, \bibinfo {author} {\bibfnamefont {Z.}~\bibnamefont {Hu}}, \bibinfo
		{author} {\bibfnamefont {B.~C.}\ \bibnamefont {Sanders}},\ and\ \bibinfo
		{author} {\bibfnamefont {S.}~\bibnamefont {Kais}},\ }\bibfield  {title}
	{\bibinfo {title} {Qudits and high-dimensional quantum computing},\ }\href
	{https://doi.org/10.3389/fphy.2020.589504} {\bibfield  {journal} {\bibinfo
			{journal} {Frontiers in Physics}\ }\textbf {\bibinfo {volume} {8}},\ \bibinfo
		{pages} {479} (\bibinfo {year} {2020})}\BibitemShut {NoStop}%
	\bibitem [{\citenamefont {Imany}\ \emph {et~al.}(2019)\citenamefont {Imany},
		\citenamefont {Jaramillo-Villegas}, \citenamefont {Alshaykh}, \citenamefont
		{Lukens}, \citenamefont {Odele}, \citenamefont {Moore}, \citenamefont
		{Leaird}, \citenamefont {Qi},\ and\ \citenamefont {Weiner}}]{Imany}%
	\BibitemOpen
	\bibfield  {author} {\bibinfo {author} {\bibfnamefont {P.}~\bibnamefont
			{Imany}}, \bibinfo {author} {\bibfnamefont {J.~A.}\ \bibnamefont
			{Jaramillo-Villegas}}, \bibinfo {author} {\bibfnamefont {M.~S.}\ \bibnamefont
			{Alshaykh}}, \bibinfo {author} {\bibfnamefont {J.~M.}\ \bibnamefont
			{Lukens}}, \bibinfo {author} {\bibfnamefont {O.~D.}\ \bibnamefont {Odele}},
		\bibinfo {author} {\bibfnamefont {A.~J.}\ \bibnamefont {Moore}}, \bibinfo
		{author} {\bibfnamefont {D.~E.}\ \bibnamefont {Leaird}}, \bibinfo {author}
		{\bibfnamefont {M.}~\bibnamefont {Qi}},\ and\ \bibinfo {author}
		{\bibfnamefont {A.~M.}\ \bibnamefont {Weiner}},\ }\bibfield  {title}
	{\bibinfo {title} {High-dimensional optical quantum logic in large
			operational spaces},\ }\href {https://doi.org/10.1038/s41534-019-0173-8}
	{\bibfield  {journal} {\bibinfo  {journal} {npj Quantum Inf.}\ }\textbf
		{\bibinfo {volume} {5}},\ \bibinfo {pages} {59} (\bibinfo {year}
		{2019})}\BibitemShut {NoStop}%
	\bibitem [{\citenamefont {Kiktenko}\ \emph
		{et~al.}(2015{\natexlab{a}})\citenamefont {Kiktenko}, \citenamefont
		{Fedorov}, \citenamefont {Man'ko},\ and\ \citenamefont
		{Man'ko}}]{PhysRevA.91.042312}%
	\BibitemOpen
	\bibfield  {author} {\bibinfo {author} {\bibfnamefont {E.~O.}\ \bibnamefont
			{Kiktenko}}, \bibinfo {author} {\bibfnamefont {A.~K.}\ \bibnamefont
			{Fedorov}}, \bibinfo {author} {\bibfnamefont {O.~V.}\ \bibnamefont
			{Man'ko}},\ and\ \bibinfo {author} {\bibfnamefont {V.~I.}\ \bibnamefont
			{Man'ko}},\ }\bibfield  {title} {\bibinfo {title} {Multilevel superconducting
			circuits as two-qubit systems: Operations, state preparation, and entropic
			inequalities},\ }\href {https://doi.org/10.1103/PhysRevA.91.042312}
	{\bibfield  {journal} {\bibinfo  {journal} {Phys. Rev. A}\ }\textbf {\bibinfo
			{volume} {91}},\ \bibinfo {pages} {042312} (\bibinfo {year}
		{2015}{\natexlab{a}})}\BibitemShut {NoStop}%
	\bibitem [{\citenamefont {Mischuck}\ and\ \citenamefont
		{M\o{}lmer}(2013)}]{PhysRevA.87.022341}%
	\BibitemOpen
	\bibfield  {author} {\bibinfo {author} {\bibfnamefont {B.}~\bibnamefont
			{Mischuck}}\ and\ \bibinfo {author} {\bibfnamefont {K.}~\bibnamefont
			{M\o{}lmer}},\ }\bibfield  {title} {\bibinfo {title} {Qudit quantum
			computation in the {J}aynes-{C}ummings model},\ }\href
	{https://doi.org/10.1103/PhysRevA.87.022341} {\bibfield  {journal} {\bibinfo
			{journal} {Phys. Rev. A}\ }\textbf {\bibinfo {volume} {87}},\ \bibinfo
		{pages} {022341} (\bibinfo {year} {2013})}\BibitemShut {NoStop}%
	\bibitem [{\citenamefont {Lanyon}\ \emph {et~al.}(2009)\citenamefont {Lanyon},
		\citenamefont {Barbieri}, \citenamefont {Almeida}, \citenamefont {Jennewein},
		\citenamefont {Ralph}, \citenamefont {Resch}, \citenamefont {Pryde},
		\citenamefont {Jeremy L. O'Brien~and},\ and\ \citenamefont
		{White}}]{NatPhysToffoli}%
	\BibitemOpen
	\bibfield  {author} {\bibinfo {author} {\bibfnamefont {B.~P.}\ \bibnamefont
			{Lanyon}}, \bibinfo {author} {\bibfnamefont {M.}~\bibnamefont {Barbieri}},
		\bibinfo {author} {\bibfnamefont {M.~P.}\ \bibnamefont {Almeida}}, \bibinfo
		{author} {\bibfnamefont {T.}~\bibnamefont {Jennewein}}, \bibinfo {author}
		{\bibfnamefont {T.~C.}\ \bibnamefont {Ralph}}, \bibinfo {author}
		{\bibfnamefont {K.~J.}\ \bibnamefont {Resch}}, \bibinfo {author}
		{\bibfnamefont {G.~J.}\ \bibnamefont {Pryde}}, \bibinfo {author}
		{\bibfnamefont {A.~G.}\ \bibnamefont {Jeremy L. O'Brien~and}},\ and\ \bibinfo
		{author} {\bibfnamefont {A.~G.}\ \bibnamefont {White}},\ }\bibfield  {title}
	{\bibinfo {title} {Simplifying quantum logic using higher-dimensional
			{H}ilbert spaces},\ }\href {https://doi.org/10.1038/nphys1150} {\bibfield
		{journal} {\bibinfo  {journal} {Nature Phys.}\ }\textbf {\bibinfo {volume}
			{5}},\ \bibinfo {pages} {134} (\bibinfo {year} {2009})}\BibitemShut {NoStop}%
	\bibitem [{\citenamefont {Kiktenko}\ \emph
		{et~al.}(2015{\natexlab{b}})\citenamefont {Kiktenko}, \citenamefont
		{Fedorov}, \citenamefont {Strakhov},\ and\ \citenamefont
		{Man'ko}}]{DeutschQudit}%
	\BibitemOpen
	\bibfield  {author} {\bibinfo {author} {\bibfnamefont {E.}~\bibnamefont
			{Kiktenko}}, \bibinfo {author} {\bibfnamefont {A.}~\bibnamefont {Fedorov}},
		\bibinfo {author} {\bibfnamefont {A.}~\bibnamefont {Strakhov}},\ and\
		\bibinfo {author} {\bibfnamefont {V.}~\bibnamefont {Man'ko}},\ }\bibfield
	{title} {\bibinfo {title} {Single qudit realization of the {D}eutsch
			algorithm using superconducting many-level quantum circuits},\ }\href
	{https://doi.org/https://doi.org/10.1016/j.physleta.2015.03.023} {\bibfield
		{journal} {\bibinfo  {journal} {Physics Letters A}\ }\textbf {\bibinfo
			{volume} {379}},\ \bibinfo {pages} {1409 } (\bibinfo {year}
		{2015}{\natexlab{b}})}\BibitemShut {NoStop}%
	\bibitem [{\citenamefont {Godfrin}\ \emph {et~al.}(2017)\citenamefont
		{Godfrin}, \citenamefont {Ferhat}, \citenamefont {Ballou}, \citenamefont
		{Klyatskaya}, \citenamefont {Ruben}, \citenamefont {Wernsdorfer},\ and\
		\citenamefont {Balestro}}]{GroverWW}%
	\BibitemOpen
	\bibfield  {author} {\bibinfo {author} {\bibfnamefont {C.}~\bibnamefont
			{Godfrin}}, \bibinfo {author} {\bibfnamefont {A.}~\bibnamefont {Ferhat}},
		\bibinfo {author} {\bibfnamefont {R.}~\bibnamefont {Ballou}}, \bibinfo
		{author} {\bibfnamefont {S.}~\bibnamefont {Klyatskaya}}, \bibinfo {author}
		{\bibfnamefont {M.}~\bibnamefont {Ruben}}, \bibinfo {author} {\bibfnamefont
			{W.}~\bibnamefont {Wernsdorfer}},\ and\ \bibinfo {author} {\bibfnamefont
			{F.}~\bibnamefont {Balestro}},\ }\bibfield  {title} {\bibinfo {title}
		{Operating quantum states in single magnetic molecules: Implementation of
			{G}rover's quantum algorithm.},\ }\href
	{https://doi.org/https://doi.org/10.1103/PhysRevLett.119.187702} {\bibfield
		{journal} {\bibinfo  {journal} {Phys. Rev. Lett.}\ }\textbf {\bibinfo
			{volume} {119}},\ \bibinfo {pages} {187702} (\bibinfo {year}
		{2017})}\BibitemShut {NoStop}%
	\bibitem [{\citenamefont {Tacchino}\ \emph {et~al.}()\citenamefont {Tacchino},
		\citenamefont {Chiesa}, \citenamefont {Sessoli}, \citenamefont {Tavernelli},\
		and\ \citenamefont {Carretta}}]{TacchinoQudits}%
	\BibitemOpen
	\bibfield  {author} {\bibinfo {author} {\bibfnamefont {F.}~\bibnamefont
			{Tacchino}}, \bibinfo {author} {\bibfnamefont {A.}~\bibnamefont {Chiesa}},
		\bibinfo {author} {\bibfnamefont {R.}~\bibnamefont {Sessoli}}, \bibinfo
		{author} {\bibfnamefont {I.}~\bibnamefont {Tavernelli}},\ and\ \bibinfo
		{author} {\bibfnamefont {S.}~\bibnamefont {Carretta}},\ }\bibfield  {title}
	{\bibinfo {title} {Molecular spin qudits for quantum simulation of
			light-matter interactions},\ }\href@noop {} {\bibinfo  {journal} {submitted}\
	}\BibitemShut {NoStop}%
	\bibitem [{\citenamefont {Michael}\ \emph {et~al.}(2016)\citenamefont
		{Michael}, \citenamefont {Silveri}, \citenamefont {Brierley}, \citenamefont
		{Albert}, \citenamefont {Salmilehto}, \citenamefont {Jiang},\ and\
		\citenamefont {Girvin}}]{PRXGirvin}%
	\BibitemOpen
	\bibfield  {journal} {  }\bibfield  {author} {\bibinfo {author} {\bibfnamefont
			{M.~H.}\ \bibnamefont {Michael}}, \bibinfo {author} {\bibfnamefont
			{M.}~\bibnamefont {Silveri}}, \bibinfo {author} {\bibfnamefont
			{R.}~\bibnamefont {Brierley}}, \bibinfo {author} {\bibfnamefont {V.~V.}\
			\bibnamefont {Albert}}, \bibinfo {author} {\bibfnamefont {J.}~\bibnamefont
			{Salmilehto}}, \bibinfo {author} {\bibfnamefont {L.}~\bibnamefont {Jiang}},\
		and\ \bibinfo {author} {\bibfnamefont {S.~M.}\ \bibnamefont {Girvin}},\
	}\bibfield  {title} {\bibinfo {title} {New class of quantum error-correcting
			codes for a bosonic mode},\ }\href
	{https://doi.org/https://doi.org/10.1103/PhysRevX.6.031006} {\bibfield
		{journal} {\bibinfo  {journal} {Phys. Rev. X}\ }\textbf {\bibinfo {volume}
			{6}},\ \bibinfo {pages} {031006} (\bibinfo {year} {2016})}\BibitemShut
	{NoStop}%
	\bibitem [{\citenamefont {Cafaro}\ \emph {et~al.}(2012)\citenamefont {Cafaro},
		\citenamefont {Maiolini},\ and\ \citenamefont {Mancini}}]{Mancini}%
	\BibitemOpen
	\bibfield  {author} {\bibinfo {author} {\bibfnamefont {C.}~\bibnamefont
			{Cafaro}}, \bibinfo {author} {\bibfnamefont {F.}~\bibnamefont {Maiolini}},\
		and\ \bibinfo {author} {\bibfnamefont {S.}~\bibnamefont {Mancini}},\
	}\bibfield  {title} {\bibinfo {title} {Quantum stabilizer codes embedding
			qubits into qudits},\ }\href {https://doi.org/10.1103/PhysRevA.86.022308}
	{\bibfield  {journal} {\bibinfo  {journal} {Phys. Rev. A}\ }\textbf {\bibinfo
			{volume} {86}},\ \bibinfo {pages} {022308} (\bibinfo {year}
		{2012})}\BibitemShut {NoStop}%
	\bibitem [{\citenamefont {Albert}\ \emph {et~al.}(2020)\citenamefont {Albert},
		\citenamefont {Covey},\ and\ \citenamefont {Preskill}}]{PRXPreskill}%
	\BibitemOpen
	\bibfield  {author} {\bibinfo {author} {\bibfnamefont {V.~V.}\ \bibnamefont
			{Albert}}, \bibinfo {author} {\bibfnamefont {J.~P.}\ \bibnamefont {Covey}},\
		and\ \bibinfo {author} {\bibfnamefont {J.}~\bibnamefont {Preskill}},\
	}\bibfield  {title} {\bibinfo {title} {Robust encoding of a qubit in a
			molecule},\ }\href {https://doi.org/10.1103/PhysRevX.10.031050} {\bibfield
		{journal} {\bibinfo  {journal} {Phys. Rev. X}\ }\textbf {\bibinfo {volume}
			{10}},\ \bibinfo {pages} {031050} (\bibinfo {year} {2020})}\BibitemShut
	{NoStop}%
	\bibitem [{\citenamefont {Hussain}\ \emph {et~al.}(2018)\citenamefont
		{Hussain}, \citenamefont {Allodi}, \citenamefont {Chiesa}, \citenamefont
		{Garlatti}, \citenamefont {Mitcov}, \citenamefont {Konstantatos},
		\citenamefont {Pedersen}, \citenamefont {Renzi}, \citenamefont {Piligkos},\
		and\ \citenamefont {Carretta}}]{jacsYb}%
	\BibitemOpen
	\bibfield  {author} {\bibinfo {author} {\bibfnamefont {R.}~\bibnamefont
			{Hussain}}, \bibinfo {author} {\bibfnamefont {G.}~\bibnamefont {Allodi}},
		\bibinfo {author} {\bibfnamefont {A.}~\bibnamefont {Chiesa}}, \bibinfo
		{author} {\bibfnamefont {E.}~\bibnamefont {Garlatti}}, \bibinfo {author}
		{\bibfnamefont {D.}~\bibnamefont {Mitcov}}, \bibinfo {author} {\bibfnamefont
			{A.}~\bibnamefont {Konstantatos}}, \bibinfo {author} {\bibfnamefont
			{K.}~\bibnamefont {Pedersen}}, \bibinfo {author} {\bibfnamefont {R.~D.}\
			\bibnamefont {Renzi}}, \bibinfo {author} {\bibfnamefont {S.}~\bibnamefont
			{Piligkos}},\ and\ \bibinfo {author} {\bibfnamefont {S.}~\bibnamefont
			{Carretta}},\ }\bibfield  {title} {\bibinfo {title} {Coherent manipulation of
			a molecular ln-based nuclear qudit coupled to an electron qubit},\ }\href
	{https://doi.org/https://doi.org/10.1021/jacs.8b05934} {\bibfield  {journal}
		{\bibinfo  {journal} {J. Am. Chem. Soc.}\ }\textbf {\bibinfo {volume}
			{140}},\ \bibinfo {pages} {9814} (\bibinfo {year} {2018})}\BibitemShut
	{NoStop}%
	\bibitem [{\citenamefont {Chiesa}\ \emph {et~al.}(2020)\citenamefont {Chiesa},
		\citenamefont {Macaluso}, \citenamefont {Petiziol}, \citenamefont
		{Wimberger}, \citenamefont {Santini},\ and\ \citenamefont
		{Carretta}}]{JPCLqec}%
	\BibitemOpen
	\bibfield  {author} {\bibinfo {author} {\bibfnamefont {A.}~\bibnamefont
			{Chiesa}}, \bibinfo {author} {\bibfnamefont {E.}~\bibnamefont {Macaluso}},
		\bibinfo {author} {\bibfnamefont {F.}~\bibnamefont {Petiziol}}, \bibinfo
		{author} {\bibfnamefont {S.}~\bibnamefont {Wimberger}}, \bibinfo {author}
		{\bibfnamefont {P.}~\bibnamefont {Santini}},\ and\ \bibinfo {author}
		{\bibfnamefont {S.}~\bibnamefont {Carretta}},\ }\bibfield  {title} {\bibinfo
		{title} {Molecular nanomagnets as qubits with embedded quantum-error
			correction},\ }\href
	{https://doi.org/https://doi.org/10.1021/acs.jpclett.0c02213} {\bibfield
		{journal} {\bibinfo  {journal} {J. Phys. Chem. Lett.}\ }\textbf {\bibinfo
			{volume} {11}},\ \bibinfo {pages} {8610} (\bibinfo {year}
		{2020})}\BibitemShut {NoStop}%
	\bibitem [{\citenamefont {Macaluso}\ \emph {et~al.}(2020)\citenamefont
		{Macaluso}, \citenamefont {Rub\'in}, \citenamefont {Aguil\`{a}},
		\citenamefont {Chiesa}, \citenamefont {L.~A.~Barrios}, \citenamefont
		{Alonso}, \citenamefont {Roubeau}, \citenamefont {Luis}, \citenamefont
		{Arom\'i},\ and\ \citenamefont {Carretta}}]{ErCeEr}%
	\BibitemOpen
	\bibfield  {author} {\bibinfo {author} {\bibfnamefont {E.}~\bibnamefont
			{Macaluso}}, \bibinfo {author} {\bibfnamefont {M.}~\bibnamefont {Rub\'in}},
		\bibinfo {author} {\bibfnamefont {D.}~\bibnamefont {Aguil\`{a}}}, \bibinfo
		{author} {\bibfnamefont {A.}~\bibnamefont {Chiesa}}, \bibinfo {author}
		{\bibfnamefont {J.~I.~M.}\ \bibnamefont {L.~A.~Barrios}}, \bibinfo {author}
		{\bibfnamefont {P.~J.}\ \bibnamefont {Alonso}}, \bibinfo {author}
		{\bibfnamefont {O.}~\bibnamefont {Roubeau}}, \bibinfo {author} {\bibfnamefont
			{F.}~\bibnamefont {Luis}}, \bibinfo {author} {\bibfnamefont {G.}~\bibnamefont
			{Arom\'i}},\ and\ \bibinfo {author} {\bibfnamefont {S.}~\bibnamefont
			{Carretta}},\ }\bibfield  {title} {\bibinfo {title} {A heterometallic
			[lnln'ln] lanthanide complex as a qubit with embedded quantum error
			correction},\ }\href {https://doi.org/10.1039/D0SC03107K} {\bibfield
		{journal} {\bibinfo  {journal} {Chem. Sci.}\ }\textbf {\bibinfo {volume}
			{11}},\ \bibinfo {pages} {10337} (\bibinfo {year} {2020})}\BibitemShut
	{NoStop}%
	\bibitem [{\citenamefont {Gaita-Ari$\tilde{\text n}$o}\ \emph
		{et~al.}(2019)\citenamefont {Gaita-Ari$\tilde{\text n}$o}, \citenamefont
		{Luis}, \citenamefont {Hill},\ and\ \citenamefont {Coronado}}]{Gaitarev}%
	\BibitemOpen
	\bibfield  {author} {\bibinfo {author} {\bibfnamefont {A.}~\bibnamefont
			{Gaita-Ari$\tilde{\text n}$o}}, \bibinfo {author} {\bibfnamefont
			{F.}~\bibnamefont {Luis}}, \bibinfo {author} {\bibfnamefont {S.}~\bibnamefont
			{Hill}},\ and\ \bibinfo {author} {\bibfnamefont {E.}~\bibnamefont
			{Coronado}},\ }\bibfield  {title} {\bibinfo {title} {Molecular spins for
			quantum computation.},\ }\href
	{https://doi.org/https://doi.org/10.1038/s41557-019-0232-y} {\bibfield
		{journal} {\bibinfo  {journal} {Nature Chem.}\ }\textbf {\bibinfo {volume}
			{11}},\ \bibinfo {pages} {301} (\bibinfo {year} {2019})}\BibitemShut
	{NoStop}%
	\bibitem [{\citenamefont {Atzori}\ and\ \citenamefont
		{Sessoli}(2019)}]{Sessoli2019}%
	\BibitemOpen
	\bibfield  {author} {\bibinfo {author} {\bibfnamefont {M.}~\bibnamefont
			{Atzori}}\ and\ \bibinfo {author} {\bibfnamefont {R.}~\bibnamefont
			{Sessoli}},\ }\bibfield  {title} {\bibinfo {title} {The second quantum
			revolution: Role and challenges of molecular chemistry},\ }\href
	{https://doi.org/https://doi.org/10.1021/jacs.9b00984} {\bibfield  {journal}
		{\bibinfo  {journal} {J. Am. Chem. Soc.}\ }\textbf {\bibinfo {volume}
			{141}},\ \bibinfo {pages} {11339} (\bibinfo {year} {2019})}\BibitemShut
	{NoStop}%
	\bibitem [{\citenamefont {Ako}\ \emph {et~al.}(2006)\citenamefont {Ako},
		\citenamefont {Hewitt}, \citenamefont {Mereacre}, \citenamefont {Cl\'{e}rac},
		\citenamefont {Wernsdorfer}, \citenamefont {Anson},\ and\ \citenamefont
		{Powell}}]{Mn19powell}%
	\BibitemOpen
	\bibfield  {author} {\bibinfo {author} {\bibfnamefont {A.~M.}\ \bibnamefont
			{Ako}}, \bibinfo {author} {\bibfnamefont {I.~J.}\ \bibnamefont {Hewitt}},
		\bibinfo {author} {\bibfnamefont {V.}~\bibnamefont {Mereacre}}, \bibinfo
		{author} {\bibfnamefont {R.}~\bibnamefont {Cl\'{e}rac}}, \bibinfo {author}
		{\bibfnamefont {W.}~\bibnamefont {Wernsdorfer}}, \bibinfo {author}
		{\bibfnamefont {C.~E.}\ \bibnamefont {Anson}},\ and\ \bibinfo {author}
		{\bibfnamefont {A.~K.}\ \bibnamefont {Powell}},\ }\bibfield  {title}
	{\bibinfo {title} {A ferromagnetically coupled {M}n19 aggregate with a record
			s=83/2 ground spin state},\ }\href@noop {} {\bibfield  {journal} {\bibinfo
			{journal} {Angewandte Chemie International Edition}\ }\textbf {\bibinfo
			{volume} {45}},\ \bibinfo {pages} {4926} (\bibinfo {year}
		{2006})}\BibitemShut {NoStop}%
	\bibitem [{\citenamefont {Baniodeh}\ \emph {et~al.}(2018)\citenamefont
		{Baniodeh}, \citenamefont {Magnani}, \citenamefont {Lan}, \citenamefont
		{Buth}, \citenamefont {Anson}, \citenamefont {Richter}, \citenamefont
		{Affronte}, \citenamefont {Schnack},\ and\ \citenamefont {Powell}}]{giant}%
	\BibitemOpen
	\bibfield  {author} {\bibinfo {author} {\bibfnamefont {A.}~\bibnamefont
			{Baniodeh}}, \bibinfo {author} {\bibfnamefont {N.}~\bibnamefont {Magnani}},
		\bibinfo {author} {\bibfnamefont {Y.}~\bibnamefont {Lan}}, \bibinfo {author}
		{\bibfnamefont {G.}~\bibnamefont {Buth}}, \bibinfo {author} {\bibfnamefont
			{C.~E.}\ \bibnamefont {Anson}}, \bibinfo {author} {\bibfnamefont
			{J.}~\bibnamefont {Richter}}, \bibinfo {author} {\bibfnamefont
			{M.}~\bibnamefont {Affronte}}, \bibinfo {author} {\bibfnamefont
			{J.}~\bibnamefont {Schnack}},\ and\ \bibinfo {author} {\bibfnamefont {A.~K.}\
			\bibnamefont {Powell}},\ }\bibfield  {title} {\bibinfo {title} {High spin
			cycles: topping the spin record for a single molecule verging on quantum
			criticality.},\ }\href
	{https://doi.org/https://doi.org/10.1038/s41535-018-0082-7} {\bibfield
		{journal} {\bibinfo  {journal} {npj Quant. Mater.}\ }\textbf {\bibinfo
			{volume} {3}},\ \bibinfo {pages} {10} (\bibinfo {year} {2018})}\BibitemShut
	{NoStop}%
	\bibitem [{\citenamefont {Wedge}\ \emph {et~al.}(2012)\citenamefont {Wedge},
		\citenamefont {Timco}, \citenamefont {Spielberg}, \citenamefont {George},
		\citenamefont {Tuna}, \citenamefont {Rigby}, \citenamefont {McInnes},
		\citenamefont {P.Winpenny}, \citenamefont {Blundell},\ and\ \citenamefont
		{Ardavan}}]{PRLWedge}%
	\BibitemOpen
	\bibfield  {author} {\bibinfo {author} {\bibfnamefont {C.~J.}\ \bibnamefont
			{Wedge}}, \bibinfo {author} {\bibfnamefont {G.~A.}\ \bibnamefont {Timco}},
		\bibinfo {author} {\bibfnamefont {E.~T.}\ \bibnamefont {Spielberg}}, \bibinfo
		{author} {\bibfnamefont {R.~E.}\ \bibnamefont {George}}, \bibinfo {author}
		{\bibfnamefont {F.}~\bibnamefont {Tuna}}, \bibinfo {author} {\bibfnamefont
			{S.}~\bibnamefont {Rigby}}, \bibinfo {author} {\bibfnamefont {E.~J.~L.}\
			\bibnamefont {McInnes}}, \bibinfo {author} {\bibfnamefont {R.~E.}\
			\bibnamefont {P.Winpenny}}, \bibinfo {author} {\bibfnamefont {S.~J.}\
			\bibnamefont {Blundell}},\ and\ \bibinfo {author} {\bibfnamefont
			{A.}~\bibnamefont {Ardavan}},\ }\bibfield  {title} {\bibinfo {title}
		{Chemical engineering of molecular qubits.},\ }\href
	{https://doi.org/https://doi.org/10.1103/PhysRevLett.108.107204} {\bibfield
		{journal} {\bibinfo  {journal} {Phys. Rev. Lett.}\ }\textbf {\bibinfo
			{volume} {108}},\ \bibinfo {pages} {107204} (\bibinfo {year}
		{2012})}\BibitemShut {NoStop}%
	\bibitem [{\citenamefont {Bader}\ \emph {et~al.}(2014)\citenamefont {Bader},
		\citenamefont {Dengler}, \citenamefont {Lenz}, \citenamefont {Endeward},
		\citenamefont {Jiang}, \citenamefont {Neugebauer},\ and\ \citenamefont {van
			Slageren}}]{Bader}%
	\BibitemOpen
	\bibfield  {author} {\bibinfo {author} {\bibfnamefont {K.}~\bibnamefont
			{Bader}}, \bibinfo {author} {\bibfnamefont {D.}~\bibnamefont {Dengler}},
		\bibinfo {author} {\bibfnamefont {S.}~\bibnamefont {Lenz}}, \bibinfo {author}
		{\bibfnamefont {B.}~\bibnamefont {Endeward}}, \bibinfo {author}
		{\bibfnamefont {S.-D.}\ \bibnamefont {Jiang}}, \bibinfo {author}
		{\bibfnamefont {P.}~\bibnamefont {Neugebauer}},\ and\ \bibinfo {author}
		{\bibfnamefont {J.}~\bibnamefont {van Slageren}},\ }\bibfield  {title}
	{\bibinfo {title} {Room temperature quantum coherence in a potential
			molecular qubit},\ }\href
	{https://doi.org/https://doi.org/10.1038/ncomms6304} {\bibfield  {journal}
		{\bibinfo  {journal} {Nat. Commun}\ }\textbf {\bibinfo {volume} {5}},\
		\bibinfo {pages} {5304} (\bibinfo {year} {2014})}\BibitemShut {NoStop}%
	\bibitem [{\citenamefont {Zadrozny}\ \emph {et~al.}(2015)\citenamefont
		{Zadrozny}, \citenamefont {Niklas}, \citenamefont {Poluektov},\ and\
		\citenamefont {Freedman}}]{Zadrozny}%
	\BibitemOpen
	\bibfield  {author} {\bibinfo {author} {\bibfnamefont {J.~M.}\ \bibnamefont
			{Zadrozny}}, \bibinfo {author} {\bibfnamefont {J.}~\bibnamefont {Niklas}},
		\bibinfo {author} {\bibfnamefont {O.~G.}\ \bibnamefont {Poluektov}},\ and\
		\bibinfo {author} {\bibfnamefont {D.~E.}\ \bibnamefont {Freedman}},\
	}\bibfield  {title} {\bibinfo {title} {Millisecond coherence time in a
			tunable molecular electronic spin qubit.},\ }\href
	{https://doi.org/https://doi.org/10.1021/acscentsci.5b00338} {\bibfield
		{journal} {\bibinfo  {journal} {ACS Cent. Sci.}\ }\textbf {\bibinfo {volume}
			{1}},\ \bibinfo {pages} {488} (\bibinfo {year} {2015})}\BibitemShut {NoStop}%
	\bibitem [{\citenamefont {Atzori}\ \emph
		{et~al.}(2016{\natexlab{a}})\citenamefont {Atzori}, \citenamefont {Tesi},
		\citenamefont {Morra}, \citenamefont {Chiesa}, \citenamefont {Sorace},\ and\
		\citenamefont {Sessoli}}]{Atzori2016}%
	\BibitemOpen
	\bibfield  {author} {\bibinfo {author} {\bibfnamefont {M.}~\bibnamefont
			{Atzori}}, \bibinfo {author} {\bibfnamefont {L.}~\bibnamefont {Tesi}},
		\bibinfo {author} {\bibfnamefont {E.}~\bibnamefont {Morra}}, \bibinfo
		{author} {\bibfnamefont {M.}~\bibnamefont {Chiesa}}, \bibinfo {author}
		{\bibfnamefont {L.}~\bibnamefont {Sorace}},\ and\ \bibinfo {author}
		{\bibfnamefont {R.}~\bibnamefont {Sessoli}},\ }\bibfield  {title} {\bibinfo
		{title} {Room-temperature quantum coherence and rabi oscillations in vanadyl
			phthalocyanine: Toward multifunctional molecular spin qubits.},\ }\href
	{https://doi.org/https://doi.org/10.1021/jacs.5b13408} {\bibfield  {journal}
		{\bibinfo  {journal} {J. Am. Chem. Soc.}\ }\textbf {\bibinfo {volume}
			{138}},\ \bibinfo {pages} {2154} (\bibinfo {year}
		{2016}{\natexlab{a}})}\BibitemShut {NoStop}%
	\bibitem [{\citenamefont {Atzori}\ \emph
		{et~al.}(2016{\natexlab{b}})\citenamefont {Atzori}, \citenamefont {Morra},
		\citenamefont {Tesi}, \citenamefont {Albino}, \citenamefont {Chiesa},
		\citenamefont {Sorace},\ and\ \citenamefont {Sessoli}}]{Atzori_JACS}%
	\BibitemOpen
	\bibfield  {author} {\bibinfo {author} {\bibfnamefont {M.}~\bibnamefont
			{Atzori}}, \bibinfo {author} {\bibfnamefont {E.}~\bibnamefont {Morra}},
		\bibinfo {author} {\bibfnamefont {L.}~\bibnamefont {Tesi}}, \bibinfo {author}
		{\bibfnamefont {A.}~\bibnamefont {Albino}}, \bibinfo {author} {\bibfnamefont
			{M.}~\bibnamefont {Chiesa}}, \bibinfo {author} {\bibfnamefont
			{L.}~\bibnamefont {Sorace}},\ and\ \bibinfo {author} {\bibfnamefont
			{R.}~\bibnamefont {Sessoli}},\ }\bibfield  {title} {\bibinfo {title} {Quantum
			coherence times enhancement in vanadium(iv)-based potential molecular qubits:
			the key role of the vanadyl moiety},\ }\href
	{https://doi.org/https://doi.org/10.1021/jacs.6b05574} {\bibfield  {journal}
		{\bibinfo  {journal} {J. Am. Chem. Soc.}\ }\textbf {\bibinfo {volume}
			{138}},\ \bibinfo {pages} {11234} (\bibinfo {year}
		{2016}{\natexlab{b}})}\BibitemShut {NoStop}%
	\bibitem [{\citenamefont {Atzori}\ \emph {et~al.}(2017)\citenamefont {Atzori},
		\citenamefont {Tesi}, \citenamefont {Benci}, \citenamefont {Lunghi},
		\citenamefont {Righini}, \citenamefont {Taschin}, \citenamefont {Torre},
		\citenamefont {Sorace},\ and\ \citenamefont {Sessoli}}]{Atzori2017}%
	\BibitemOpen
	\bibfield  {author} {\bibinfo {author} {\bibfnamefont {M.}~\bibnamefont
			{Atzori}}, \bibinfo {author} {\bibfnamefont {L.}~\bibnamefont {Tesi}},
		\bibinfo {author} {\bibfnamefont {S.}~\bibnamefont {Benci}}, \bibinfo
		{author} {\bibfnamefont {A.}~\bibnamefont {Lunghi}}, \bibinfo {author}
		{\bibfnamefont {R.}~\bibnamefont {Righini}}, \bibinfo {author} {\bibfnamefont
			{A.}~\bibnamefont {Taschin}}, \bibinfo {author} {\bibfnamefont
			{R.}~\bibnamefont {Torre}}, \bibinfo {author} {\bibfnamefont
			{L.}~\bibnamefont {Sorace}},\ and\ \bibinfo {author} {\bibfnamefont
			{R.}~\bibnamefont {Sessoli}},\ }\bibfield  {title} {\bibinfo {title} {Spin
			dynamics and low energy vibrations: Insights from vanadyl- based potential
			molecular qubits.},\ }\href
	{https://doi.org/https://pubs.acs.org/doi/abs/10.1021/jacs.6b05574}
	{\bibfield  {journal} {\bibinfo  {journal} {J. Am. Chem. Soc.}\ }\textbf
		{\bibinfo {volume} {139}},\ \bibinfo {pages} {4338} (\bibinfo {year}
		{2017})}\BibitemShut {NoStop}%
	\bibitem [{\citenamefont {Atzori}\ \emph
		{et~al.}(2018{\natexlab{a}})\citenamefont {Atzori}, \citenamefont {Benci},
		\citenamefont {Morra}, \citenamefont {Tesi}, \citenamefont {Chiesa},
		\citenamefont {Torre}, \citenamefont {Sorace},\ and\ \citenamefont
		{Sessoli}}]{Atzori2018}%
	\BibitemOpen
	\bibfield  {author} {\bibinfo {author} {\bibfnamefont {M.}~\bibnamefont
			{Atzori}}, \bibinfo {author} {\bibfnamefont {S.}~\bibnamefont {Benci}},
		\bibinfo {author} {\bibfnamefont {E.}~\bibnamefont {Morra}}, \bibinfo
		{author} {\bibfnamefont {L.}~\bibnamefont {Tesi}}, \bibinfo {author}
		{\bibfnamefont {M.}~\bibnamefont {Chiesa}}, \bibinfo {author} {\bibfnamefont
			{R.}~\bibnamefont {Torre}}, \bibinfo {author} {\bibfnamefont
			{L.}~\bibnamefont {Sorace}},\ and\ \bibinfo {author} {\bibfnamefont
			{R.}~\bibnamefont {Sessoli}},\ }\bibfield  {title} {\bibinfo {title}
		{Structural effects on the spin dynamics of potential molecular qubits.},\
	}\href {https://doi.org/https://doi.org/10.1021/acs.inorgchem.7b02616}
	{\bibfield  {journal} {\bibinfo  {journal} {Inorg. Chem.}\ }\textbf {\bibinfo
			{volume} {57}},\ \bibinfo {pages} {731} (\bibinfo {year}
		{2018}{\natexlab{a}})}\BibitemShut {NoStop}%
	\bibitem [{\citenamefont {Yu}\ \emph {et~al.}(2016)\citenamefont {Yu},
		\citenamefont {Graham}, \citenamefont {Zadrozny}, \citenamefont {Niklas},
		\citenamefont {Krzyaniak}, \citenamefont {Wasielewski}, \citenamefont
		{Poluektov},\ and\ \citenamefont {Freedman}}]{Freedman_JACS}%
	\BibitemOpen
	\bibfield  {author} {\bibinfo {author} {\bibfnamefont {C.-J.}\ \bibnamefont
			{Yu}}, \bibinfo {author} {\bibfnamefont {M.~J.}\ \bibnamefont {Graham}},
		\bibinfo {author} {\bibfnamefont {J.~M.}\ \bibnamefont {Zadrozny}}, \bibinfo
		{author} {\bibfnamefont {J.}~\bibnamefont {Niklas}}, \bibinfo {author}
		{\bibfnamefont {M.~D.}\ \bibnamefont {Krzyaniak}}, \bibinfo {author}
		{\bibfnamefont {M.~R.}\ \bibnamefont {Wasielewski}}, \bibinfo {author}
		{\bibfnamefont {O.~G.}\ \bibnamefont {Poluektov}},\ and\ \bibinfo {author}
		{\bibfnamefont {D.~E.}\ \bibnamefont {Freedman}},\ }\bibfield  {title}
	{\bibinfo {title} {Long coherence times in nuclear spin-free vanadyl
			qubits},\ }\href {https://doi.org/https://doi.org/10.1021/jacs.6b08467}
	{\bibfield  {journal} {\bibinfo  {journal} {J. Am. Chem. Soc.}\ }\textbf
		{\bibinfo {volume} {138}},\ \bibinfo {pages} {14678} (\bibinfo {year}
		{2016})}\BibitemShut {NoStop}%
	\bibitem [{\citenamefont {Fataftah}\ \emph {et~al.}(2016)\citenamefont
		{Fataftah}, \citenamefont {Zadrozny}, \citenamefont {Coste}, \citenamefont
		{Graham}, \citenamefont {Rogers},\ and\ \citenamefont
		{Freedman}}]{Freedman_Cr}%
	\BibitemOpen
	\bibfield  {author} {\bibinfo {author} {\bibfnamefont {M.}~\bibnamefont
			{Fataftah}}, \bibinfo {author} {\bibfnamefont {J.~M.}\ \bibnamefont
			{Zadrozny}}, \bibinfo {author} {\bibfnamefont {S.~C.}\ \bibnamefont {Coste}},
		\bibinfo {author} {\bibfnamefont {M.~J.}\ \bibnamefont {Graham}}, \bibinfo
		{author} {\bibfnamefont {D.~M.}\ \bibnamefont {Rogers}},\ and\ \bibinfo
		{author} {\bibfnamefont {D.~E.}\ \bibnamefont {Freedman}},\ }\bibfield
	{title} {\bibinfo {title} {Employing forbidden transitions as qubits in a
			nuclear spin-free chromium complex.},\ }\href
	{https://doi.org/https://doi.org/10.1021/jacs.5b11802} {\bibfield  {journal}
		{\bibinfo  {journal} {J. Am. Chem. Soc.}\ }\textbf {\bibinfo {volume}
			{138}},\ \bibinfo {pages} {1344} (\bibinfo {year} {2016})}\BibitemShut
	{NoStop}%
	\bibitem [{\citenamefont {Graham}\ \emph {et~al.}(2014)\citenamefont {Graham},
		\citenamefont {Zadrozny}, \citenamefont {Shiddiq}, \citenamefont {Anderson},
		\citenamefont {Fataftah}, \citenamefont {Hill},\ and\ \citenamefont
		{Freedman}}]{Freedman2014}%
	\BibitemOpen
	\bibfield  {author} {\bibinfo {author} {\bibfnamefont {M.~J.}\ \bibnamefont
			{Graham}}, \bibinfo {author} {\bibfnamefont {J.~M.}\ \bibnamefont
			{Zadrozny}}, \bibinfo {author} {\bibfnamefont {M.}~\bibnamefont {Shiddiq}},
		\bibinfo {author} {\bibfnamefont {J.~S.}\ \bibnamefont {Anderson}}, \bibinfo
		{author} {\bibfnamefont {M.~S.}\ \bibnamefont {Fataftah}}, \bibinfo {author}
		{\bibfnamefont {S.}~\bibnamefont {Hill}},\ and\ \bibinfo {author}
		{\bibfnamefont {D.~E.}\ \bibnamefont {Freedman}},\ }\bibfield  {title}
	{\bibinfo {title} {Influence of electronic spin and spin–orbit coupling on
			decoherence in mononuclear transition metal complexes.},\ }\href
	{https://doi.org/https://pubs.acs.org/doi/abs/10.1021/acs.inorgchem.5b02429}
	{\bibfield  {journal} {\bibinfo  {journal} {J. Am. Chem. Soc.}\ }\textbf
		{\bibinfo {volume} {136}},\ \bibinfo {pages} {7623} (\bibinfo {year}
		{2014})}\BibitemShut {NoStop}%
	\bibitem [{\citenamefont {Zadrozny}\ and\ \citenamefont
		{Freedman}(2015)}]{Zadrozny_Fe}%
	\BibitemOpen
	\bibfield  {author} {\bibinfo {author} {\bibfnamefont {J.}~\bibnamefont
			{Zadrozny}}\ and\ \bibinfo {author} {\bibfnamefont {D.~E.}\ \bibnamefont
			{Freedman}},\ }\bibfield  {title} {\bibinfo {title} {Qubit control limited by
			spin–lattice relaxation in a nuclear spin-free iron (iii) complex.},\
	}\href {https://doi.org/https://doi.org/10.1021/acs.inorgchem.5b02429}
	{\bibfield  {journal} {\bibinfo  {journal} {Inorg. Chem.}\ }\textbf {\bibinfo
			{volume} {54}},\ \bibinfo {pages} {12027} (\bibinfo {year}
		{2015})}\BibitemShut {NoStop}%
	\bibitem [{\citenamefont {Luis}\ \emph {et~al.}(2011)\citenamefont {Luis},
		\citenamefont {Repoll\'es}, \citenamefont {Mart\'inez-P\'erez}, \citenamefont
		{Aguil\'a}, \citenamefont {Roubeau}, \citenamefont {Zueco}, \citenamefont
		{Alonso}, \citenamefont {Evangelisti}, \citenamefont {A.~Cam\'on},
		\citenamefont {Barrios},\ and\ \citenamefont {Arom\'i}}]{Luis2011}%
	\BibitemOpen
	\bibfield  {author} {\bibinfo {author} {\bibfnamefont {F.}~\bibnamefont
			{Luis}}, \bibinfo {author} {\bibfnamefont {A.}~\bibnamefont {Repoll\'es}},
		\bibinfo {author} {\bibfnamefont {M.~J.}\ \bibnamefont {Mart\'inez-P\'erez}},
		\bibinfo {author} {\bibfnamefont {D.}~\bibnamefont {Aguil\'a}}, \bibinfo
		{author} {\bibfnamefont {O.}~\bibnamefont {Roubeau}}, \bibinfo {author}
		{\bibfnamefont {D.}~\bibnamefont {Zueco}}, \bibinfo {author} {\bibfnamefont
			{P.~J.}\ \bibnamefont {Alonso}}, \bibinfo {author} {\bibfnamefont
			{M.}~\bibnamefont {Evangelisti}}, \bibinfo {author} {\bibfnamefont {J.~S.}\
			\bibnamefont {A.~Cam\'on}}, \bibinfo {author} {\bibfnamefont {L.~A.}\
			\bibnamefont {Barrios}},\ and\ \bibinfo {author} {\bibfnamefont
			{G.}~\bibnamefont {Arom\'i}},\ }\bibfield  {title} {\bibinfo {title}
		{Molecular prototypes for spin-based cnot and swap quantum gates.},\ }\href
	{https://doi.org/https://doi.org/10.1103/PhysRevLett.107.117203} {\bibfield
		{journal} {\bibinfo  {journal} {Phys. Rev. Lett.}\ }\textbf {\bibinfo
			{volume} {107}},\ \bibinfo {pages} {117203} (\bibinfo {year}
		{2011})}\BibitemShut {NoStop}%
	\bibitem [{\citenamefont {Aguil\`{a}}\ \emph {et~al.}(2014)\citenamefont
		{Aguil\`{a}}, \citenamefont {Barrios}, \citenamefont {Velasco}, \citenamefont
		{Roubeau}, \citenamefont {Repoll\'es}, \citenamefont {Alonso}, \citenamefont
		{Ses\'e}, \citenamefont {Teat}, \citenamefont {Luis},\ and\ \citenamefont
		{Arom\'i}}]{Aguila2014}%
	\BibitemOpen
	\bibfield  {author} {\bibinfo {author} {\bibfnamefont {D.}~\bibnamefont
			{Aguil\`{a}}}, \bibinfo {author} {\bibfnamefont {D.}~\bibnamefont {Barrios}},
		\bibinfo {author} {\bibfnamefont {V.}~\bibnamefont {Velasco}}, \bibinfo
		{author} {\bibfnamefont {O.}~\bibnamefont {Roubeau}}, \bibinfo {author}
		{\bibfnamefont {A.}~\bibnamefont {Repoll\'es}}, \bibinfo {author}
		{\bibfnamefont {P.}~\bibnamefont {Alonso}}, \bibinfo {author} {\bibfnamefont
			{J.}~\bibnamefont {Ses\'e}}, \bibinfo {author} {\bibfnamefont
			{S.}~\bibnamefont {Teat}}, \bibinfo {author} {\bibfnamefont {F.}~\bibnamefont
			{Luis}},\ and\ \bibinfo {author} {\bibfnamefont {G.}~\bibnamefont
			{Arom\'i}},\ }\bibfield  {title} {\bibinfo {title} {Heterodimetallic [lnln']
			lanthanide complexes: Toward a chemical design of two-qubit molecular spin
			quantum gates.},\ }\href {https://doi.org/https://doi.org/10.1038/srep07423}
	{\bibfield  {journal} {\bibinfo  {journal} {J. Am. Chem. Soc.}\ }\textbf
		{\bibinfo {volume} {136}},\ \bibinfo {pages} {14215} (\bibinfo {year}
		{2014})}\BibitemShut {NoStop}%
	\bibitem [{\citenamefont {Ardavan}\ \emph {et~al.}(2015)\citenamefont
		{Ardavan}, \citenamefont {Bowen}, \citenamefont {Fernandez}, \citenamefont
		{Fielding}, \citenamefont {Kaminski}, \citenamefont {Moro}, \citenamefont
		{Muryn}, \citenamefont {D.Wise}, \citenamefont {Ruggi}, \citenamefont
		{McInnes}, \citenamefont {Severin}, \citenamefont {G.~A.~Timco},
		\citenamefont {Tuna}, \citenamefont {Whitehead},\ and\ \citenamefont
		{P.Winpenny}}]{Ardavan2015}%
	\BibitemOpen
	\bibfield  {author} {\bibinfo {author} {\bibfnamefont {A.}~\bibnamefont
			{Ardavan}}, \bibinfo {author} {\bibfnamefont {A.~M.}\ \bibnamefont {Bowen}},
		\bibinfo {author} {\bibfnamefont {A.}~\bibnamefont {Fernandez}}, \bibinfo
		{author} {\bibfnamefont {A.~J.}\ \bibnamefont {Fielding}}, \bibinfo {author}
		{\bibfnamefont {D.}~\bibnamefont {Kaminski}}, \bibinfo {author}
		{\bibfnamefont {F.}~\bibnamefont {Moro}}, \bibinfo {author} {\bibfnamefont
			{C.~A.}\ \bibnamefont {Muryn}}, \bibinfo {author} {\bibfnamefont
			{M.}~\bibnamefont {D.Wise}}, \bibinfo {author} {\bibfnamefont
			{A.}~\bibnamefont {Ruggi}}, \bibinfo {author} {\bibfnamefont {E.~J.~L.}\
			\bibnamefont {McInnes}}, \bibinfo {author} {\bibfnamefont {K.}~\bibnamefont
			{Severin}}, \bibinfo {author} {\bibfnamefont {C.~R.~T.}\ \bibnamefont
			{G.~A.~Timco}}, \bibinfo {author} {\bibfnamefont {F.}~\bibnamefont {Tuna}},
		\bibinfo {author} {\bibfnamefont {G.~F.~S.}\ \bibnamefont {Whitehead}},\ and\
		\bibinfo {author} {\bibfnamefont {R.~E.}\ \bibnamefont {P.Winpenny}},\
	}\bibfield  {title} {\bibinfo {title} {Engineering coherent interactions in
			molecular nanomagnet dimers.},\ }\href
	{https://doi.org/https://doi.org/10.1038/npjqi.2015.12} {\bibfield  {journal}
		{\bibinfo  {journal} {npj Quantum Information}\ }\textbf {\bibinfo {volume}
			{1}},\ \bibinfo {pages} {15012} (\bibinfo {year} {2015})}\BibitemShut
	{NoStop}%
	\bibitem [{\citenamefont {Chiesa}\ \emph {et~al.}(2014)\citenamefont {Chiesa},
		\citenamefont {Whitehead}, \citenamefont {Carretta}, \citenamefont {Carthy},
		\citenamefont {Timco}, \citenamefont {Teat}, \citenamefont {Amoretti},
		\citenamefont {Pavarini}, \citenamefont {Winpenny},\ and\ \citenamefont
		{Santini}}]{SciRepNi}%
	\BibitemOpen
	\bibfield  {author} {\bibinfo {author} {\bibfnamefont {A.}~\bibnamefont
			{Chiesa}}, \bibinfo {author} {\bibfnamefont {G.~F.~S.}\ \bibnamefont
			{Whitehead}}, \bibinfo {author} {\bibfnamefont {S.}~\bibnamefont {Carretta}},
		\bibinfo {author} {\bibfnamefont {L.}~\bibnamefont {Carthy}}, \bibinfo
		{author} {\bibfnamefont {G.~A.}\ \bibnamefont {Timco}}, \bibinfo {author}
		{\bibfnamefont {S.~J.}\ \bibnamefont {Teat}}, \bibinfo {author}
		{\bibfnamefont {G.}~\bibnamefont {Amoretti}}, \bibinfo {author}
		{\bibfnamefont {E.}~\bibnamefont {Pavarini}}, \bibinfo {author}
		{\bibfnamefont {R.~E.~P.}\ \bibnamefont {Winpenny}},\ and\ \bibinfo {author}
		{\bibfnamefont {P.}~\bibnamefont {Santini}},\ }\bibfield  {title} {\bibinfo
		{title} {Molecular nanomagnets with switchable coupling for quantum
			simulation.},\ }\href {https://doi.org/https://doi.org/10.1038/srep07423}
	{\bibfield  {journal} {\bibinfo  {journal} {Sci. Rep.}\ }\textbf {\bibinfo
			{volume} {4}},\ \bibinfo {pages} {7423} (\bibinfo {year} {2014})}\BibitemShut
	{NoStop}%
	\bibitem [{\citenamefont {Ferrando-Soria}\ \emph {et~al.}(2016)\citenamefont
		{Ferrando-Soria}, \citenamefont {Moreno-Pineda}, \citenamefont {Chiesa},
		\citenamefont {Fernandez}, \citenamefont {Magee}, \citenamefont {Carretta},
		\citenamefont {Santini}, \citenamefont {Vitorica-Yrezabal}, \citenamefont
		{Tuna}, \citenamefont {McInness},\ and\ \citenamefont {Winpenny}}]{modules}%
	\BibitemOpen
	\bibfield  {author} {\bibinfo {author} {\bibfnamefont {J.}~\bibnamefont
			{Ferrando-Soria}}, \bibinfo {author} {\bibfnamefont {E.}~\bibnamefont
			{Moreno-Pineda}}, \bibinfo {author} {\bibfnamefont {A.}~\bibnamefont
			{Chiesa}}, \bibinfo {author} {\bibfnamefont {A.}~\bibnamefont {Fernandez}},
		\bibinfo {author} {\bibfnamefont {S.~A.}\ \bibnamefont {Magee}}, \bibinfo
		{author} {\bibfnamefont {S.}~\bibnamefont {Carretta}}, \bibinfo {author}
		{\bibfnamefont {P.}~\bibnamefont {Santini}}, \bibinfo {author} {\bibfnamefont
			{I.}~\bibnamefont {Vitorica-Yrezabal}}, \bibinfo {author} {\bibfnamefont
			{F.}~\bibnamefont {Tuna}}, \bibinfo {author} {\bibfnamefont {E.~J.~L.}\
			\bibnamefont {McInness}},\ and\ \bibinfo {author} {\bibfnamefont {R.~E.~P.}\
			\bibnamefont {Winpenny}},\ }\bibfield  {title} {\bibinfo {title} {A modular
			design of molecular qubits to implement universal quantum gates.},\ }\href
	{https://doi.org/https://doi.org/10.1038/ncomms11377} {\bibfield  {journal}
		{\bibinfo  {journal} {Nat. Commun.}\ }\textbf {\bibinfo {volume} {7}},\
		\bibinfo {pages} {11377} (\bibinfo {year} {2016})}\BibitemShut {NoStop}%
	\bibitem [{\citenamefont {{Ferrando-Soria et al.}}(2016)}]{Chem}%
	\BibitemOpen
	\bibfield  {author} {\bibinfo {author} {\bibfnamefont {J.}~\bibnamefont
			{{Ferrando-Soria et al.}}},\ }\bibfield  {title} {\bibinfo {title}
		{Swithcable interactions in molecular double qubits.},\ }\href
	{https://doi.org/https://doi.org/10.1016/j.chempr.2016.10.001} {\bibfield
		{journal} {\bibinfo  {journal} {Chem}\ }\textbf {\bibinfo {volume} {1}},\
		\bibinfo {pages} {727} (\bibinfo {year} {2016})}\BibitemShut {NoStop}%
	\bibitem [{\citenamefont {Atzori}\ \emph
		{et~al.}(2018{\natexlab{b}})\citenamefont {Atzori}, \citenamefont {Chiesa},
		\citenamefont {Morra}, \citenamefont {Chiesa}, \citenamefont {Sorace},
		\citenamefont {Carretta},\ and\ \citenamefont {Sessoli}}]{VO2}%
	\BibitemOpen
	\bibfield  {author} {\bibinfo {author} {\bibfnamefont {M.}~\bibnamefont
			{Atzori}}, \bibinfo {author} {\bibfnamefont {A.}~\bibnamefont {Chiesa}},
		\bibinfo {author} {\bibfnamefont {E.}~\bibnamefont {Morra}}, \bibinfo
		{author} {\bibfnamefont {M.}~\bibnamefont {Chiesa}}, \bibinfo {author}
		{\bibfnamefont {L.}~\bibnamefont {Sorace}}, \bibinfo {author} {\bibfnamefont
			{S.}~\bibnamefont {Carretta}},\ and\ \bibinfo {author} {\bibfnamefont
			{R.}~\bibnamefont {Sessoli}},\ }\bibfield  {title} {\bibinfo {title} {A
			two-qubit molecular architecture for electronmediated nuclear quantum
			simulation.},\ }\href {https://doi.org/10.1039/C8SC01695J} {\bibfield
		{journal} {\bibinfo  {journal} {Chem. Sci.}\ }\textbf {\bibinfo {volume}
			{9}},\ \bibinfo {pages} {6183} (\bibinfo {year}
		{2018}{\natexlab{b}})}\BibitemShut {NoStop}%
	\bibitem [{\citenamefont {Garlatti}\ \emph {et~al.}(2021)\citenamefont
		{Garlatti}, \citenamefont {Guidi}, \citenamefont {Ansbro}, \citenamefont
		{Santini}, \citenamefont {Amoretti}, \citenamefont {Ollivier}, \citenamefont
		{Mutka}, \citenamefont {Timco}, \citenamefont {Vitorica-Yrezabal},
		\citenamefont {Whitehead}, \citenamefont {Winpenny},\ and\ \citenamefont
		{Carretta}}]{Entanglement}%
	\BibitemOpen
	\bibfield  {author} {\bibinfo {author} {\bibfnamefont {E.}~\bibnamefont
			{Garlatti}}, \bibinfo {author} {\bibfnamefont {T.}~\bibnamefont {Guidi}},
		\bibinfo {author} {\bibfnamefont {S.}~\bibnamefont {Ansbro}}, \bibinfo
		{author} {\bibfnamefont {P.}~\bibnamefont {Santini}}, \bibinfo {author}
		{\bibfnamefont {G.}~\bibnamefont {Amoretti}}, \bibinfo {author}
		{\bibfnamefont {J.}~\bibnamefont {Ollivier}}, \bibinfo {author}
		{\bibfnamefont {H.}~\bibnamefont {Mutka}}, \bibinfo {author} {\bibfnamefont
			{G.}~\bibnamefont {Timco}}, \bibinfo {author} {\bibfnamefont {I.~J.}\
			\bibnamefont {Vitorica-Yrezabal}}, \bibinfo {author} {\bibfnamefont
			{G.~F.~S.}\ \bibnamefont {Whitehead}}, \bibinfo {author} {\bibfnamefont
			{R.~E.~P.}\ \bibnamefont {Winpenny}},\ and\ \bibinfo {author} {\bibfnamefont
			{S.}~\bibnamefont {Carretta}},\ }\bibfield  {title} {\bibinfo {title}
		{Portraying entanglement between molecular qubits with four-dimensional
			inelastic neutron scattering},\ }\bibfield  {journal} {\bibinfo  {journal}
		{Nature Communications}\ }\textbf {\bibinfo {volume} {8}},\ \href
	{https://doi.org/https://doi.org/10.1038/ncomms14543}
	{https://doi.org/10.1038/ncomms14543} (\bibinfo {year} {2021})\BibitemShut
	{NoStop}%
	\bibitem [{\citenamefont {Timco}\ \emph {et~al.}(2016)\citenamefont {Timco},
		\citenamefont {Marocchi}, \citenamefont {Garlatti}, \citenamefont {Barker},
		\citenamefont {Albring}, \citenamefont {Bellini}, \citenamefont {Manghi},
		\citenamefont {McInnes}, \citenamefont {Pritchard}, \citenamefont {Tuna},
		\citenamefont {Wernsdorfer}, \citenamefont {Lorusso}, \citenamefont
		{Amoretti}, \citenamefont {Carretta}, \citenamefont {Affronte},\ and\
		\citenamefont {Winpenny}}]{Heterodimers}%
	\BibitemOpen
	\bibfield  {author} {\bibinfo {author} {\bibfnamefont {G.}~\bibnamefont
			{Timco}}, \bibinfo {author} {\bibfnamefont {S.}~\bibnamefont {Marocchi}},
		\bibinfo {author} {\bibfnamefont {E.}~\bibnamefont {Garlatti}}, \bibinfo
		{author} {\bibfnamefont {C.}~\bibnamefont {Barker}}, \bibinfo {author}
		{\bibfnamefont {M.}~\bibnamefont {Albring}}, \bibinfo {author} {\bibfnamefont
			{V.}~\bibnamefont {Bellini}}, \bibinfo {author} {\bibfnamefont
			{F.}~\bibnamefont {Manghi}}, \bibinfo {author} {\bibfnamefont {E.~J.~L.}\
			\bibnamefont {McInnes}}, \bibinfo {author} {\bibfnamefont {R.~G.}\
			\bibnamefont {Pritchard}}, \bibinfo {author} {\bibfnamefont {F.}~\bibnamefont
			{Tuna}}, \bibinfo {author} {\bibfnamefont {W.}~\bibnamefont {Wernsdorfer}},
		\bibinfo {author} {\bibfnamefont {G.}~\bibnamefont {Lorusso}}, \bibinfo
		{author} {\bibfnamefont {G.}~\bibnamefont {Amoretti}}, \bibinfo {author}
		{\bibfnamefont {S.}~\bibnamefont {Carretta}}, \bibinfo {author}
		{\bibfnamefont {M.}~\bibnamefont {Affronte}},\ and\ \bibinfo {author}
		{\bibfnamefont {R.~E.~P.}\ \bibnamefont {Winpenny}},\ }\bibfield  {title}
	{\bibinfo {title} {Heterodimers of heterometallic rings},\ }\href
	{https://doi.org/10.1039/C6DT01941B} {\bibfield  {journal} {\bibinfo
			{journal} {Dalton Trans.}\ }\textbf {\bibinfo {volume} {45}},\ \bibinfo
		{pages} {16610} (\bibinfo {year} {2016})}\BibitemShut {NoStop}%
	\bibitem [{\citenamefont {Zhang}\ \emph {et~al.}(2015)\citenamefont {Zhang},
		\citenamefont {Burgarth}, \citenamefont {Laflamme},\ and\ \citenamefont
		{Suter}}]{Actuator}%
	\BibitemOpen
	\bibfield  {author} {\bibinfo {author} {\bibfnamefont {J.}~\bibnamefont
			{Zhang}}, \bibinfo {author} {\bibfnamefont {D.}~\bibnamefont {Burgarth}},
		\bibinfo {author} {\bibfnamefont {R.}~\bibnamefont {Laflamme}},\ and\
		\bibinfo {author} {\bibfnamefont {D.}~\bibnamefont {Suter}},\ }\bibfield
	{title} {\bibinfo {title} {Experimental implementation of quantum gates
			through actuator qubits},\ }\href
	{https://doi.org/10.1103/PhysRevA.91.012330} {\bibfield  {journal} {\bibinfo
			{journal} {Phys. Rev. A}\ }\textbf {\bibinfo {volume} {91}},\ \bibinfo
		{pages} {012330} (\bibinfo {year} {2015})}\BibitemShut {NoStop}%
	\bibitem [{\citenamefont {Troiani}\ and\ \citenamefont
		{Paris}(2016)}]{PRBTroiani2016}%
	\BibitemOpen
	\bibfield  {author} {\bibinfo {author} {\bibfnamefont {F.}~\bibnamefont
			{Troiani}}\ and\ \bibinfo {author} {\bibfnamefont {M.~G.~A.}\ \bibnamefont
			{Paris}},\ }\bibfield  {title} {\bibinfo {title} {Probing molecular spin
			clusters by local measurements},\ }\href
	{https://doi.org/10.1103/PhysRevB.94.115422} {\bibfield  {journal} {\bibinfo
			{journal} {Phys. Rev. B}\ }\textbf {\bibinfo {volume} {94}},\ \bibinfo
		{pages} {115422} (\bibinfo {year} {2016})}\BibitemShut {NoStop}%
	\bibitem [{\citenamefont {Ghirardi}\ \emph {et~al.}(2018)\citenamefont
		{Ghirardi}, \citenamefont {Siloi}, \citenamefont {Bordone}, \citenamefont
		{Troiani},\ and\ \citenamefont {Paris}}]{PRAGhirardi2018}%
	\BibitemOpen
	\bibfield  {author} {\bibinfo {author} {\bibfnamefont {L.}~\bibnamefont
			{Ghirardi}}, \bibinfo {author} {\bibfnamefont {I.}~\bibnamefont {Siloi}},
		\bibinfo {author} {\bibfnamefont {P.}~\bibnamefont {Bordone}}, \bibinfo
		{author} {\bibfnamefont {F.}~\bibnamefont {Troiani}},\ and\ \bibinfo {author}
		{\bibfnamefont {M.~G.~A.}\ \bibnamefont {Paris}},\ }\bibfield  {title}
	{\bibinfo {title} {Quantum metrology at level anticrossing},\ }\href
	{https://doi.org/10.1103/PhysRevA.97.012120} {\bibfield  {journal} {\bibinfo
			{journal} {Phys. Rev. A}\ }\textbf {\bibinfo {volume} {97}},\ \bibinfo
		{pages} {012120} (\bibinfo {year} {2018})}\BibitemShut {NoStop}%
	\bibitem [{\citenamefont {Yamabayashi}\ \emph {et~al.}(2018)\citenamefont
		{Yamabayashi}, \citenamefont {Atzori}, \citenamefont {L.~Tesi}, \citenamefont
		{Santanni}, \citenamefont {Boulon}, \citenamefont {Morra}, \citenamefont
		{Benci}, \citenamefont {Torre}, \citenamefont {Chiesa}, \citenamefont
		{Sorace}, \citenamefont {Sessoli},\ and\ \citenamefont {Yamashita}}]{VOTPP}%
	\BibitemOpen
	\bibfield  {author} {\bibinfo {author} {\bibfnamefont {T.}~\bibnamefont
			{Yamabayashi}}, \bibinfo {author} {\bibfnamefont {M.}~\bibnamefont {Atzori}},
		\bibinfo {author} {\bibfnamefont {G.}~\bibnamefont {L.~Tesi}}, \bibinfo
		{author} {\bibfnamefont {F.}~\bibnamefont {Santanni}}, \bibinfo {author}
		{\bibfnamefont {M.}~\bibnamefont {Boulon}}, \bibinfo {author} {\bibfnamefont
			{E.}~\bibnamefont {Morra}}, \bibinfo {author} {\bibfnamefont
			{S.}~\bibnamefont {Benci}}, \bibinfo {author} {\bibfnamefont
			{R.}~\bibnamefont {Torre}}, \bibinfo {author} {\bibfnamefont
			{M.}~\bibnamefont {Chiesa}}, \bibinfo {author} {\bibfnamefont
			{L.}~\bibnamefont {Sorace}}, \bibinfo {author} {\bibfnamefont
			{R.}~\bibnamefont {Sessoli}},\ and\ \bibinfo {author} {\bibfnamefont
			{M.}~\bibnamefont {Yamashita}},\ }\bibfield  {title} {\bibinfo {title}
		{Scaling up electronic spin qubits into a three-dimensional metal−organic
			framework},\ }\href {https://doi.org/https://doi.org/10.1021/jacs.8b06733}
	{\bibfield  {journal} {\bibinfo  {journal} {J. Am. Chem. Soc.}\ }\textbf
		{\bibinfo {volume} {140}},\ \bibinfo {pages} {12090−12101} (\bibinfo {year}
		{2018})}\BibitemShut {NoStop}%
	\bibitem [{\citenamefont {Bonizzoni}\ \emph {et~al.}(2020)\citenamefont
		{Bonizzoni}, \citenamefont {Ghirri}, \citenamefont {Santanni}, \citenamefont
		{Atzori}, \citenamefont {Sorace}, \citenamefont {Sessoli},\ and\
		\citenamefont {Affronte}}]{Bonizzoni}%
	\BibitemOpen
	\bibfield  {author} {\bibinfo {author} {\bibfnamefont {C.}~\bibnamefont
			{Bonizzoni}}, \bibinfo {author} {\bibfnamefont {A.}~\bibnamefont {Ghirri}},
		\bibinfo {author} {\bibfnamefont {F.}~\bibnamefont {Santanni}}, \bibinfo
		{author} {\bibfnamefont {M.}~\bibnamefont {Atzori}}, \bibinfo {author}
		{\bibfnamefont {L.}~\bibnamefont {Sorace}}, \bibinfo {author} {\bibfnamefont
			{R.}~\bibnamefont {Sessoli}},\ and\ \bibinfo {author} {\bibfnamefont
			{M.}~\bibnamefont {Affronte}},\ }\bibfield  {title} {\bibinfo {title}
		{Storage and retrieval of microwave pulses with molecular spin ensembles},\
	}\href@noop {} {\bibfield  {journal} {\bibinfo  {journal} {npj Quantum
				Information}\ }\textbf {\bibinfo {volume} {6}},\ \bibinfo {pages} {68}
		(\bibinfo {year} {2020})}\BibitemShut {NoStop}%
	\bibitem [{\citenamefont {Gimeno}\ \emph {et~al.}(2021)\citenamefont {Gimeno},
		\citenamefont {Urtizberea}, \citenamefont {Román-Roche}, \citenamefont
		{Zueco}, \citenamefont {Camón}, \citenamefont {Alonso}, \citenamefont
		{Roubeau},\ and\ \citenamefont {Luis}}]{gimeno2021broadband}%
	\BibitemOpen
	\bibfield  {author} {\bibinfo {author} {\bibfnamefont {I.}~\bibnamefont
			{Gimeno}}, \bibinfo {author} {\bibfnamefont {A.}~\bibnamefont {Urtizberea}},
		\bibinfo {author} {\bibfnamefont {J.}~\bibnamefont {Román-Roche}}, \bibinfo
		{author} {\bibfnamefont {D.}~\bibnamefont {Zueco}}, \bibinfo {author}
		{\bibfnamefont {A.}~\bibnamefont {Camón}}, \bibinfo {author} {\bibfnamefont
			{P.~J.}\ \bibnamefont {Alonso}}, \bibinfo {author} {\bibfnamefont
			{O.}~\bibnamefont {Roubeau}},\ and\ \bibinfo {author} {\bibfnamefont
			{F.}~\bibnamefont {Luis}},\ }\href@noop {} {\bibinfo {title} {Broad-band
			spectroscopy of a vanadyl porphyrin: a model electronuclear spin qudit}}
	(\bibinfo {year} {2021}),\ \Eprint {https://arxiv.org/abs/2101.11650}
	{arXiv:2101.11650 [quant-ph]} \BibitemShut {NoStop}%
	\bibitem [{\citenamefont {Garlatti}\ \emph {et~al.}(2020)\citenamefont
		{Garlatti}, \citenamefont {Tesi}, \citenamefont {Lunghi}, \citenamefont
		{Atzori}, \citenamefont {Voneshen}, \citenamefont {Santini}, \citenamefont
		{Sanvito}, \citenamefont {Guidi}, \citenamefont {Sessoli},\ and\
		\citenamefont {Carretta}}]{Phonons}%
	\BibitemOpen
	\bibfield  {author} {\bibinfo {author} {\bibfnamefont {E.}~\bibnamefont
			{Garlatti}}, \bibinfo {author} {\bibfnamefont {L.}~\bibnamefont {Tesi}},
		\bibinfo {author} {\bibfnamefont {A.}~\bibnamefont {Lunghi}}, \bibinfo
		{author} {\bibfnamefont {M.}~\bibnamefont {Atzori}}, \bibinfo {author}
		{\bibfnamefont {D.~J.}\ \bibnamefont {Voneshen}}, \bibinfo {author}
		{\bibfnamefont {P.}~\bibnamefont {Santini}}, \bibinfo {author} {\bibfnamefont
			{S.}~\bibnamefont {Sanvito}}, \bibinfo {author} {\bibfnamefont
			{T.}~\bibnamefont {Guidi}}, \bibinfo {author} {\bibfnamefont
			{R.}~\bibnamefont {Sessoli}},\ and\ \bibinfo {author} {\bibfnamefont
			{S.}~\bibnamefont {Carretta}},\ }\bibfield  {title} {\bibinfo {title}
		{Unveiling phonons in a molecular qubit with four-dimensional inelastic
			neutron scattering and density functional theory},\ }\bibfield  {journal}
	{\bibinfo  {journal} {Nature Communications}\ }\textbf {\bibinfo {volume}
		{11}},\ \href {https://doi.org/10.1038/s41467-020-15475-7}
	{10.1038/s41467-020-15475-7} (\bibinfo {year} {2020})\BibitemShut {NoStop}%
	\bibitem [{\citenamefont {Feng}\ \emph {et~al.}()\citenamefont {Feng},
		\citenamefont {Gu}, \citenamefont {Li}, \citenamefont {Jiang}, \citenamefont
		{Wei},\ and\ \citenamefont {Zhou}}]{Synthesis}%
	\BibitemOpen
	\bibfield  {author} {\bibinfo {author} {\bibfnamefont {D.}~\bibnamefont
			{Feng}}, \bibinfo {author} {\bibfnamefont {Z.-Y.}\ \bibnamefont {Gu}},
		\bibinfo {author} {\bibfnamefont {J.-R.}\ \bibnamefont {Li}}, \bibinfo
		{author} {\bibfnamefont {H.-L.}\ \bibnamefont {Jiang}}, \bibinfo {author}
		{\bibfnamefont {Z.}~\bibnamefont {Wei}},\ and\ \bibinfo {author}
		{\bibfnamefont {H.-C.}\ \bibnamefont {Zhou}},\ }\bibfield  {title} {\bibinfo
		{title} {Zirconium-metalloporphyrin pcn-222: Mesoporous metal–organic
			frameworks with ultrahigh stability as biomimetic catalysts},\ }\href
	{https://doi.org/https://doi.org/10.1002/anie.201204475} {\bibfield
		{journal} {\bibinfo  {journal} {Angewandte Chemie International Edition}\
		}\textbf {\bibinfo {volume} {51}},\ \bibinfo {pages} {10307}}\BibitemShut
	{NoStop}%
	\bibitem [{\citenamefont {Allodi}\ \emph {et~al.}(2005)\citenamefont {Allodi},
		\citenamefont {Banderini}, \citenamefont {De~Renzi},\ and\ \citenamefont
		{Vignali}}]{Spectrometer}%
	\BibitemOpen
	\bibfield  {author} {\bibinfo {author} {\bibfnamefont {G.}~\bibnamefont
			{Allodi}}, \bibinfo {author} {\bibfnamefont {A.}~\bibnamefont {Banderini}},
		\bibinfo {author} {\bibfnamefont {R.}~\bibnamefont {De~Renzi}},\ and\
		\bibinfo {author} {\bibfnamefont {C.}~\bibnamefont {Vignali}},\ }\bibfield
	{title} {\bibinfo {title} {Hyrespect: A broadband fast-averaging spectrometer
			for nuclear magnetic resonance of magnetic materials},\ }\href
	{https://doi.org/10.1063/1.2009868} {\bibfield  {journal} {\bibinfo
			{journal} {Review of Scientific Instruments}\ }\textbf {\bibinfo {volume}
			{76}},\ \bibinfo {pages} {083911} (\bibinfo {year} {2005})},\ \Eprint
	{https://arxiv.org/abs/https://doi.org/10.1063/1.2009868}
	{https://doi.org/10.1063/1.2009868} \BibitemShut {NoStop}%
	\bibitem [{\citenamefont {Stone}(2014)}]{TableG}%
	\BibitemOpen
	\bibfield  {author} {\bibinfo {author} {\bibfnamefont {N.}~\bibnamefont
			{Stone}},\ }\bibfield  {title} {\bibinfo {title} {Table of nuclear magnetic
			dipole and electric quadrupole moments},\ }\href
	{https://www-nds.iaea.org/publications/indc/indc-nds-0658.pdf} {\bibfield
		{journal} {\bibinfo  {journal} {International Atomic Energy Agency,
				INDC(NDS)-0658}\ }\textbf {\bibinfo {volume} {45-11}} (\bibinfo {year}
		{2014})}\BibitemShut {NoStop}%
	\bibitem [{\citenamefont {Hahn}(1950)}]{Han}%
	\BibitemOpen
	\bibfield  {author} {\bibinfo {author} {\bibfnamefont {E.~L.}\ \bibnamefont
			{Hahn}},\ }\bibfield  {title} {\bibinfo {title} {Spin echoes},\ }\href
	{https://doi.org/https://doi.org/10.1103/PhysRev.80.580} {\bibfield
		{journal} {\bibinfo  {journal} {Physical Review}\ }\textbf {\bibinfo {volume}
			{80}},\ \bibinfo {pages} {580} (\bibinfo {year} {1950})}\BibitemShut
	{NoStop}%
	\bibitem [{\citenamefont {Gottfried}(2015)}]{SurfaceGOTTFRIED}%
	\BibitemOpen
	\bibfield  {author} {\bibinfo {author} {\bibfnamefont {J.~M.}\ \bibnamefont
			{Gottfried}},\ }\bibfield  {title} {\bibinfo {title} {Surface chemistry of
			porphyrins and phthalocyanines},\ }\href
	{https://doi.org/https://doi.org/10.1016/j.surfrep.2015.04.001} {\bibfield
		{journal} {\bibinfo  {journal} {Surface Science Reports}\ }\textbf {\bibinfo
			{volume} {70}},\ \bibinfo {pages} {259} (\bibinfo {year} {2015})}\BibitemShut
	{NoStop}%
	\bibitem [{\citenamefont {Malavolti}\ \emph {et~al.}(2018)\citenamefont
		{Malavolti}, \citenamefont {Briganti}, \citenamefont {Hänze}, \citenamefont
		{Serrano}, \citenamefont {Cimatti}, \citenamefont {McMurtrie}, \citenamefont
		{Otero}, \citenamefont {Ohresser}, \citenamefont {Totti}, \citenamefont
		{Mannini}, \citenamefont {Sessoli},\ and\ \citenamefont {Loth}}]{FlatMol}%
	\BibitemOpen
	\bibfield  {author} {\bibinfo {author} {\bibfnamefont {L.}~\bibnamefont
			{Malavolti}}, \bibinfo {author} {\bibfnamefont {M.}~\bibnamefont {Briganti}},
		\bibinfo {author} {\bibfnamefont {M.}~\bibnamefont {Hänze}}, \bibinfo
		{author} {\bibfnamefont {G.}~\bibnamefont {Serrano}}, \bibinfo {author}
		{\bibfnamefont {I.}~\bibnamefont {Cimatti}}, \bibinfo {author} {\bibfnamefont
			{G.}~\bibnamefont {McMurtrie}}, \bibinfo {author} {\bibfnamefont
			{E.}~\bibnamefont {Otero}}, \bibinfo {author} {\bibfnamefont
			{P.}~\bibnamefont {Ohresser}}, \bibinfo {author} {\bibfnamefont
			{F.}~\bibnamefont {Totti}}, \bibinfo {author} {\bibfnamefont
			{M.}~\bibnamefont {Mannini}}, \bibinfo {author} {\bibfnamefont
			{R.}~\bibnamefont {Sessoli}},\ and\ \bibinfo {author} {\bibfnamefont
			{S.}~\bibnamefont {Loth}},\ }\bibfield  {title} {\bibinfo {title} {Tunable
			spin–superconductor coupling of spin 1/2 vanadyl phthalocyanine
			molecules},\ }\href {https://doi.org/10.1021/acs.nanolett.8b03921} {\bibfield
		{journal} {\bibinfo  {journal} {Nano Letters}\ }\textbf {\bibinfo {volume}
			{18}},\ \bibinfo {pages} {7955} (\bibinfo {year} {2018})}\BibitemShut
	{NoStop}%
	\bibitem [{\citenamefont {Zhang}\ \emph {et~al.}(2021)\citenamefont {Zhang},
		\citenamefont {Wolf}, \citenamefont {Wang}, \citenamefont {Aubin},
		\citenamefont {Bilgeri}, \citenamefont {Willke}, \citenamefont {Heinrich},\
		and\ \citenamefont {Choi}}]{EPRScanningProbe}%
	\BibitemOpen
	\bibfield  {author} {\bibinfo {author} {\bibfnamefont {X.}~\bibnamefont
			{Zhang}}, \bibinfo {author} {\bibfnamefont {C.}~\bibnamefont {Wolf}},
		\bibinfo {author} {\bibfnamefont {Y.}~\bibnamefont {Wang}}, \bibinfo {author}
		{\bibfnamefont {H.}~\bibnamefont {Aubin}}, \bibinfo {author} {\bibfnamefont
			{T.}~\bibnamefont {Bilgeri}}, \bibinfo {author} {\bibfnamefont
			{P.}~\bibnamefont {Willke}}, \bibinfo {author} {\bibfnamefont
			{A.}~\bibnamefont {Heinrich}},\ and\ \bibinfo {author} {\bibfnamefont
			{T.}~\bibnamefont {Choi}},\ }\bibfield  {title} {\bibinfo {title} {Electron
			spin resonance of single molecules and magnetic interaction through
			ligands},\ }\bibfield  {journal} {\bibinfo  {journal} {Submitted and
			available at Research Square}\ }\href
	{https://doi.org/https://doi.org/10.21203/rs.3.rs-134144/v1}
	{https://doi.org/10.21203/rs.3.rs-134144/v1} (\bibinfo {year}
	{2021})\BibitemShut {NoStop}%
	\bibitem [{\citenamefont {Wili}\ \emph {et~al.}(2019)\citenamefont {Wili},
		\citenamefont {Richert}, \citenamefont {Limburg}, \citenamefont {Clarke},
		\citenamefont {Anderson}, \citenamefont {Timmel},\ and\ \citenamefont
		{Jeschke}}]{IonsInteract}%
	\BibitemOpen
	\bibfield  {author} {\bibinfo {author} {\bibfnamefont {N.}~\bibnamefont
			{Wili}}, \bibinfo {author} {\bibfnamefont {S.}~\bibnamefont {Richert}},
		\bibinfo {author} {\bibfnamefont {B.}~\bibnamefont {Limburg}}, \bibinfo
		{author} {\bibfnamefont {S.~J.}\ \bibnamefont {Clarke}}, \bibinfo {author}
		{\bibfnamefont {H.~L.}\ \bibnamefont {Anderson}}, \bibinfo {author}
		{\bibfnamefont {C.~R.}\ \bibnamefont {Timmel}},\ and\ \bibinfo {author}
		{\bibfnamefont {G.}~\bibnamefont {Jeschke}},\ }\bibfield  {title} {\bibinfo
		{title} {Eldor-detected nmr beyond hyperfine couplings: a case study with
			cu(ii)-porphyrin dimers},\ }\href {https://doi.org/10.1039/C9CP01760G}
	{\bibfield  {journal} {\bibinfo  {journal} {Phys. Chem. Chem. Phys.}\
		}\textbf {\bibinfo {volume} {21}},\ \bibinfo {pages} {11676} (\bibinfo {year}
		{2019})}\BibitemShut {NoStop}%
	\bibitem [{\citenamefont {Degen}\ \emph {et~al.}(2017)\citenamefont {Degen},
		\citenamefont {Reinhard},\ and\ \citenamefont {Cappellaro}}]{RMPSensing}%
	\BibitemOpen
	\bibfield  {author} {\bibinfo {author} {\bibfnamefont {C.~L.}\ \bibnamefont
			{Degen}}, \bibinfo {author} {\bibfnamefont {F.}~\bibnamefont {Reinhard}},\
		and\ \bibinfo {author} {\bibfnamefont {P.}~\bibnamefont {Cappellaro}},\
	}\bibfield  {title} {\bibinfo {title} {Quantum sensing},\ }\href
	{https://doi.org/10.1103/RevModPhys.89.035002} {\bibfield  {journal}
		{\bibinfo  {journal} {Rev. Mod. Phys.}\ }\textbf {\bibinfo {volume} {89}},\
		\bibinfo {pages} {035002} (\bibinfo {year} {2017})}\BibitemShut {NoStop}%
\end{thebibliography}

%

\end{document}